\documentclass[11pt,a4paper]{article}%
\usepackage{amsmath}
\usepackage{amsfonts}
\usepackage{amssymb}
\usepackage{graphicx}
\usepackage{color}%
\providecommand{\U}[1]{\protect\rule{.1in}{.1in}}
\pdfoutput=1
\providecommand{\U}[1]{\protect\rule{.1in}{.1in}}
\begin{document}

\title{Observational constraints to a phenomenological $f\left(  R,\nabla R\right)  $-model}
\author{R. R. Cuzinatto$^{1}$\thanks{cuzinatto@gmail.com} , C. A. M. de Melo$^{1,2}%
$\thanks{cassius.anderson@gmail.com} , L. G. Medeiros$^{3}$%
\thanks{leogmedeiros@gmail.com} , P. J. Pompeia$^{4,5}$%
\thanks{pedropjp@ifi.cta.br} }
\date{}
\maketitle

\begin{center}
$^{1}$Instituto de Ci\^{e}ncia e Tecnologia, Universidade Federal de
Alfenas.\newline Rod. Jos\'{e} Aur\'{e}lio Vilela (BR 267), Km 533,
n${{}^{\circ}} $11999, CEP 37701-970, Po\c{c}os de Caldas, MG, Brazil.\newline%
$^{2}$Instituto de F\'{\i}sica Te\'{o}rica, Universidade Estadual
Paulista.\newline Rua Bento Teobaldo Ferraz 271 Bloco II, P.O. Box 70532-2,
CEP 01156-970, S\~{a}o Paulo, SP, Brazil.\newline$^{3}$Escola de Ci\^{e}ncia e
Tecnologia, Universidade Federal do Rio Grande do Norte. \newline Campus
Universit\'{a}rio, s/n - Lagoa Nova, CEP 59078-970, Natal, RN, Brazil.\newline$^{4}%
$Inst. de Fomento e Coordena\c{c}\~{a}o Industrial, Departamento
de Ci\^{e}ncia e Tecnologia Aeroespacial. \newline Pra\c{c}a Mal. Eduardo
Gomes 50, CEP 12228-901, S\~{a}o Jos\'{e} dos Campos, SP, Brazil.
\newline$^{5}$ Instituto Tecnol\'{o}gico de Aerona\'{u}tica, Departamento
de Ci\^{e}ncia e Tecnologia Aeroespacial. \newline Pra\c{c}a Mal. Eduardo
Gomes 50, CEP 12228-900, S\~{a}o Jos\'{e} dos Campos, SP, Brazil.\newline
\end{center}

\begin{abstract}
This paper analyses the cosmological consequences of a modified theory of
gravity whose action integral is built from a linear combination of the Ricci
scalar $R$ and a quadratic term in the covariant derivative of $R$. The
resulting Friedmann equations are of the fifth-order in the Hubble function.
These equations are solved numerically for a flat space section geometry and
pressureless matter. The cosmological parameters of the higher-order model are
fit using SN Ia data and X-ray gas mass fraction in galaxy clusters. The
best-fit present-day $t_{0}$ values for the deceleration parameter, jerk and
snap are given. The coupling constant $\beta$\ of the model is not univocally
determined by the data fit, but partially constrained by it. Density parameter
$\Omega_{m0}$ is also determined and shows weak correlation with the other
parameters. The model allows for two possible future scenarios: there may be
either a premature Big Rip or a Rebouncing event depending on the set of
values in the space of parameters. The analysis towards the past performed
with the best-fit parameters shows that the model is not able to accommodate a
matter-dominated stage required to the formation of structure.

\end{abstract}


\section{Introduction \label{sec-Intro}}

The cosmological constant is the simplest solution to the problem of the
present-day acceleration of the universe. It is consistent with all the
available observational tests, from the constraints imposed by CMB maps
\cite{WMAP Collaboration,Planck} to BAO picks \cite{SDSS Collaboration,2dFGRS
Collaboration}\ to SN Ia constraints \cite{SNLS Collaboration,Supernova Search
Team Collaboration,ESSENCE Collaboration,Union2,Union21} and the estimates of
fraction of baryons in galaxy clusters through X-ray detection
\cite{Schindler2002,Allen2008}. Nevertheless, there is an uncomfortable
feeling from the fact that $\Lambda$\ is not satisfactory explained as a
vacuum state of some field theory: the energy scale observed for $\Lambda$ and
predicted by quantum field theory badly disagree \cite{Weinberg1989}. This is
one of the reasons why alternative explanations for the present-day cosmic
acceleration have been put out since 1998, when it was first detected
\cite{Riess1998,Perlmutter1999}.

These proposals can be classified into two broad categories, each one related
to a modification of a different side of the Einstein field equations of
General Relativity \cite{Amendola2010}.

The modified matter approach introduces a negative pressure (that counteracts
gravity) through the energy momentum tensor $T_{\mu\nu}$\ on the right-hand
side of Einstein equations. Three types of models constitute this approach,
namely quintessence models
\cite{Ratra1988,Caldwell1998,Carrol1998,Hebecker2001,Gabi2007}, k-essence
solutions \cite{Chiba2000,Armendariz-Picon2001}\ and perfect fluid models
(such as the Chaplygin gas \cite{Kamenshchik2001,Bento2002}). The quintessence
and k-essence models are based on scalar fields whose potential (in the former
case) or the kinetic energy (in the later one) drive the accelerated
expansion. The perfect fluid models use exotic equations of state (for the
pressure as a function of the energy density) to generate negative pressure.
The weakness of k-essence and quintessence proposals lies in the fact that
they require a scalar field to be dominant today, when the energy scale is
low. This requirement demands the mass of such a field $\phi$\ to be extremely
small. One would then expect long range interaction of this field with
ordinary matter, which is highly constrained by local gravity experiments. The
difficulty with the perfect fluid models is the lack of an interpretation to
the exotic equations of state based on ordinary physics.

The second approach, called modified gravity, affects the left hand side of
Einstein equations \cite{Capozziello2011,Odintsov2011}. The subtypes of models in this
category include braneworld scenarios \cite{Dvali2000,Sahni2003},
scalar-tensor theories \cite{Bartolo2000,Perrotta2000}, non-riemannian
geometries \cite{FinslerDE,LyraDE2011,LyraDKP}, $f\left(  R\right)  $\ gravity
\cite{Capozziello2002,DeFelice2010,Sotiriou2010,Santos2008,Pires2010,Odintsov2006} and
$f\left(  T\right)  $\ theories \cite{Bengochea2009,Linder2010,Bamba2011}. In
the braneworld models there exists an additional dimension through which the
gravitational interaction occurs and by which the possibility of a cosmic
acceleration is accommodated. Scalar-tensor theories couple the scalar
curvature $R$\ with a scalar field $\phi$\ in the action $S$, allowing enough
degrees of freedom for an accelerated solution. Non-riemannian geometries are
another alternative way to explore the geometrization of the gravitational
interaction, but allowing new degrees of freedom describing different
geometric structures than curvature. The $f\left(  R\right)  $\ gravity
corresponds to models resulting from non-linear Lagrangian densities dependent
on $R$; $f\left(  T\right)  $\ theories build their cosmological models on
manifolds whose connection is equipped with torsion instead of curvature.

All these mechanisms of generating the present-day cosmic dynamics are
generally called dark energy models.

Our aim with the present paper is to contribute with the study of dark energy
in the branch of modified gravity. We explore the cosmological consequences of
a particular model in the context of the $f\left(  R,\nabla R\right)  $
gravity testing the model against some of the observational data available.
$f\left(  R,\nabla R\right)  $ gravity is a theory based on an action integral
whose kernel is a function of the Ricci scalar $R$ and terms involving
derivatives of $R$. Ref. \cite{Gottlober1990} is perhaps an early example of
what we call $f\left(  R,\nabla R\right)  $ gravity. A more recent example is
the (non-local) higher derivative cosmology presented in Refs.
\cite{Biswas2006,Biswas2010,Biswas2012}. These works intend to eliminate the
Big Bang singularity through a bounce produced by a theory based on an action
containing the regular Einstein-Hilbert term plus a String-spired term. This
last ingredient contains an infinite number of derivatives of $R$, a
requirement needed to keep this modified gravity theory ghost-free. References
\cite{Arkani-Hamed2002,Barvinsky2003}\ are\ $f\left(  R,\nabla R\right)
$-gravity prototypes concerned not with the past-incomplete inflationary
space-time but with the cosmological constant problem. In this regard, they
try to address a similar issue that interests us in the present work: the
cosmological acceleration nowadays.

Few years ago, we proposed a gauge theory based on the second order derivative
of the gauge potential $A$\ following Utiyama's procedure \cite{Annals}. This
led to a second field strength $G=\partial F+fAF$\ (depending on the field
strength $F$) and allowed us to obtain massive gauge fields among other
applications, such as the study of gauge invariance of the theory under $U(1)$
and $SU(N)$ groups when $\mathcal{L}=\mathcal{L}\left(  A,\partial
A,\partial^{2}A\right)  $. In a subsequent work, gravity was interpreted as a
second order gauge theory under the Lorentz group, and all possible
Lagrangians with linear and quadratic combinations of $F$ and $G$\ -- or the
Riemann tensor, its derivatives and their possible\ contractions, from the
geometrical point of view -- were built and classified within that context
\cite{EPJC}. In a third paper\ we selected one of those quadratic Lagrangians
for investigating the cosmological consequences of a theory that takes into
account $R$- and $\left(  \nabla R\right)  ^{2}$-terms
\cite{AstrophysSpaceSci}. That Lagrangian appears in the action $S$\ bellow --
Eq. (\ref{Action}) -- and can be understood as a particular example of an
$f\left(  R,\nabla R\right)  $\ gravity:
\begin{equation}
S=\int d^{4}x\sqrt{-g}\left(  \frac{R}{2\chi}+\frac{\beta}{8\chi}\nabla_{\mu
}R\nabla^{\mu}R-\mathcal{L}_{M}\right)  ~, \label{Action}%
\end{equation}
where $\chi=8\pi G$ (once we take $c=1$), $\beta$ is the coupling constant,
$\nabla_{\mu}$\ is the covariant derivative, and $\mathcal{L}_{M}$ is the
Lagrangian of the matter fields. As usual, $g=\det\left(  g_{\mu\nu}\right)
$.\footnote{The authors of Ref.\cite{Gottlober1990}\ use the same action
(\ref{Action}) proposed here but their argument is based on different
motivations.}

An important feature of $S$ is the presence of ghosts
(see \cite{Biswas2012} and references therein). However,
here this is not a serious issue because (\ref{Action}) should be considered as an
\emph{effective action} of a fundamental theory of gravity, this last one a ghost-free
theory. There are some consistent proposals of extended theories of gravity in the
literature. A good example is the String-inspired non-local higher derivative
gravitation model presented in Ref.~\cite{Biswas2010}. The ghost-free action of that model is:\begin{equation}
S=\int d^{4}x\sqrt{-g}\left(  \frac{M_{P}^{2}}{2}R+\frac{1}{2}R\mathcal{F}\left(  \square/M_{\ast}^{2}\right)  R\right)  \label{S String}\end{equation}
where $M_{P}$\ is the Planck mass ($M_{P}=1/\left(  8\pi G\right)  $, $G$ is
the gravitational constant), $M_{\ast}$ is the mass scale where the
higher-derivative terms become important and
\begin{equation}
\mathcal{F}\left(  \square/M_{\ast}^{2}\right)  ={\textstyle\sum\limits_{n=1}^{\infty}}
f_{n}\square^{n}~. \label{F String}\end{equation}
In this context, our higher-order model must be treated as a low energy
limiting case of (\ref{S String}) where the only relevant contribution of Eq.
(\ref{F String})\ for the present-day universe is the first term of the
series, i.e. $f_{1}\gg f_{n}$, $\forall n\geqslant2$. In this approximation,
Eq. (\ref{S String}) is equivalent to the action integral (\ref{Action}) of our
$f\left(  R,\nabla R\right)  $-cosmology\ up to a surface term. The equivalence
of both models leads to a relation between the
coupling constant $\beta$ and the parameter $f_{1}$ $\sim$ $M_{\ast}^{-2}$, namely
\begin{equation}
f_{1}=\frac{\beta}{32\pi G}~. \label{f1}\end{equation}

Taking the variation of (\ref{Action}) with respect to the metric tensor
$g_{\mu\nu}$ (cf. the metric formalism) gives the field equations
\begin{equation}
R_{\mu\nu}-\frac{1}{2}g_{\mu\nu}R+\beta H_{\mu\nu}=\chi T_{\mu\nu}~,
\label{EinsteinEq}%
\end{equation}
with%
\begin{align}
H_{\mu\nu}  &  =\nabla_{\mu}\nabla_{\nu}\left[  \square R\right]  +\frac{1}%
{2}\nabla_{\mu}R\nabla_{\nu}R\nonumber\\
&  -R_{\mu\nu}\square R-g_{\mu\nu}\square\left[  \square R\right]  -\frac
{1}{4}g_{\mu\nu}\nabla_{\rho}R\nabla^{\rho}R~, \label{H_mu_nu}%
\end{align}
where $\square=\nabla_{\mu}\nabla^{\mu}$ and $R_{\mu\nu}$ is the Ricci tensor.

A four dimensional spacetime with homogeneous and isotropic space section is
described by the Friedmann-Lema\^{\i}tre-Robertson-Walker (FLRW) line element%
\begin{equation}
ds^{2}=c^{2}dt^{2}-a^{2}\left(  t\right)  \left[  \frac{dr^{2}}{1-\kappa
r^{2}}+r^{2}\left(  d\theta^{2}+\sin^{2}\theta d\phi^{2}\right)  \right]  ~,
\label{ds2 FLRW}%
\end{equation}
where the curvature parameter $\kappa$\ assumes the values $+1,0,-1$ and
$a\left(  t\right)  $ is the scale parameter. Substituting the metric tensor
from Eq.(\ref{ds2 FLRW}) along with the energy-momentum tensor of a perfect
fluid in co-moving coordinates,%
\begin{equation}
T_{\mu\nu}=\left(  \rho+P\right)  \delta_{\mu}^{\;0}\delta_{\nu}^{\;0}%
-Pg_{\mu\nu} \label{T_mu_nu}%
\end{equation}
($\rho$ is the density and $P$ is the pressure of the components fulling the
universe),\ into the gravity field equations Eq.(\ref{EinsteinEq}) leads to
the two higher order Friedmann equations:%
\begin{equation}
H^{2}+\frac{\beta}{3}F_{1}=\frac{8\pi G}{3}\rho~; \label{Fried I}%
\end{equation}%
\begin{equation}
3H^{2}+2\dot{H}+\beta F_{2}=-8\pi GP~, \label{Fried II}%
\end{equation}
where%
\begin{align}
F_{1}  &  =18HH^{\left(  4\right)  }+108H^{2}\dddot{H}-18\dot{H}\dddot
{H}+9\ddot{H}^{2}+90H^{3}\ddot{H}\nonumber\\
&  +216H\dot{H}\ddot{H}-72\dot{H}^{3}+288\left(  H\dot{H}\right)
^{2}-216H^{4}\dot{H} \label{F1}%
\end{align}%
\begin{align}
F_{2}  &  =6H^{(5)}+54HH^{(4)}+138H^{2}\dddot{H}+126\dot{H}\dddot
{H}\nonumber\\
&  +81\ddot{H}^{2}+18H^{3}\ddot{H}+498H\dot{H}\ddot{H}+120\dot{H}^{3}%
-216H^{4}\dot{H} \label{F2}%
\end{align}
which are written in terms of the Hubble function $H=H(t)$,%
\begin{equation}
H\equiv\frac{\dot{a}}{a}~, \label{Hubble}%
\end{equation}
and where we have set $\kappa=0$ as is largely favored by the observational
data. The dot (such as in $\dot{a}$) means derivative with respect to the
cosmic time $t$ and for any arbitrary function of time $f=f(t)$,
\[
f^{\left(  n\right)  }\equiv\frac{d^{n}f}{dt^{n}}~,
\]
i.e., $H^{\left(  5\right)  }$ is the fifth-order derivative of the Hubble
function with respect to the cosmic time.

Eqs. (\ref{Fried I}) and (\ref{Fried II}) reduce to the usual Friedmann
equations if $\beta\rightarrow0$. From this system of modified Friedmann
equations one can obtain the equation expressing the covariant conservation of
the energy momentum tensor, $\nabla_{\nu}T^{\mu\nu}=0$, which reads%
\begin{equation}
\dot{\rho}+3H\left(  \rho+P\right)  =0~. \label{rho dot}%
\end{equation}
In Ref.\cite{AstrophysSpaceSci}, we considered simultaneously the set formed
by Eq. (\ref{rho dot}) and a combination of Eqs.(\ref{Fried I}) and
(\ref{Fried II}), namely%
\begin{equation}
\dot{H}+\frac{\beta}{2}\left(  F_{2}-F_{1}\right)  =-4\pi G\left(
\rho+P\right)  ~. \label{Fried}%
\end{equation}
The system of Eqs.(\ref{rho dot})-(\ref{Fried}) was solved perturbatively with
respect to the coupling constant $\beta$ for pressureless matter. Using the
observational data available, we have determined the free parameters of our
modified theory of gravity and the scale parameter\ as a function of the time.
The plot of $a=a\left(  t\right)  $ indicates that the decelerated regime
peculiar to the universe dominated by matter evolves naturally to an
accelerated scenario at recent times.

This makes it promising to perform a more careful analyzes of our $f\left(
R,\nabla R\right)  $\ model making a complete maximum likelihood fit to all parameters.

In this paper we will solve the system of Eqs.(\ref{rho dot})-(\ref{Fried I})
exactly, without using any approximation technique. Section \ref{sec-Model}%
\ presents the script to obtain the complete solution to the higher order
cosmological equations. Another improvement we shall do in the present work
(when comparing with Ref.\cite{AstrophysSpaceSci}) is to use the observational
data available to fully test our model. As far as we know, this is the first
time this is done for a higher derivative theory of gravity. In Section
\ref{sec-DataFit} we construct the likelihood function to be maximized in the
process of fitting the model parameters using data from SN Ia observations
\cite{Union2} and from the measurements of the mass fraction of gas in
clusters of galaxies \cite{Allen2008}. The results of this fit are given in
Section \ref{sec-Results}, where we show the present-day numerical values for
the deceleration parameter $q_{0}=q\left(  t_{0}\right)  $ \cite{Visser2004},%
\begin{equation}
q=-\frac{\ddot{a}}{aH^{2}} \label{q}%
\end{equation}
the \emph{jerk} function $j_{0}=j\left(  t_{0}\right)  $,%
\begin{equation}
j=\frac{\dddot{a}}{aH^{3}}~, \label{jerk}%
\end{equation}
and also of the \emph{snap} $s_{0}=s\left(  t_{0}\right)  $,%
\begin{equation}
s=\frac{a^{\left(  4\right)  }}{aH^{4}}~. \label{snap}%
\end{equation}
These last two functions are given in terms of higher-order time derivatives
of the scale factor and it is only natural that they show up here. In Section
\ref{sec-Behaviour} we perform an investigation of the dynamical evolution of
the model, comparing its behaviour towards the past\ with the standard
solutions for single component universes. Moreover, we show that our $f\left(
R,\nabla R\right)  $-model accommodates a future Big Rip or a Rebouncing
scenario depending on the values of the coupling constant $\beta$. The
Rebouncing event is the point where the universe attains its maximum size of
expansion and from which it begins to shrink. Our final comments about the
features of our model and its validity are given in Section
\ref{sec-Conclusion}.

\section{Higher order cosmological model \label{sec-Model}}

The solution of our higher-order model shall be obtained from the integration
of the pair of Eqs.(\ref{Fried I}) and (\ref{rho dot}). The definitions%
\begin{equation}
B\equiv H_{0}^{4}\beta~, \label{B(beta)}%
\end{equation}
and%
\begin{equation}
\Omega_{m}\left(  t\right)  \equiv\frac{\chi\rho}{3H_{0}^{2}}=\frac{\rho}%
{\rho_{c}}~, \label{Omega+m(rho)}%
\end{equation}
enables us to express Eq. (\ref{Fried I}) as%
\begin{equation}
\frac{H^{2}}{H_{0}^{2}}+\frac{B}{3H_{0}^{6}}F_{1}=\Omega_{m}\left(  t\right)
\label{Fried H B}%
\end{equation}
and the conservation of energy-momentum Eq.(\ref{rho dot}) as%
\begin{equation}
\dot{\Omega}_{m}+3H\Omega_{m}=0~, \label{Omega_m dot}%
\end{equation}
which is readly integrated:%
\begin{equation}
\Omega_{m}=\Omega_{m0}\left(  \frac{a_{0}}{a}\right)  ^{3}~.
\label{Omega_m(a)}%
\end{equation}
It is standard to take $a_{0}=1$. Moreover, substituting the solution
(\ref{Omega_m(a)}) into Eq. (\ref{Fried H B}), leads to a fifth-order
differential equation in $a$:%
\begin{equation}
\frac{1}{H_{0}^{2}}\dot{a}^{2}a^{4}+\frac{B}{H_{0}^{6}}\mathcal{F}_{1}%
=\Omega_{m0}a^{3} \label{Fried a B}%
\end{equation}
with%
\begin{align}
\mathcal{F}_{1}  &  =6a^{\left(  5\right)  }\dot{a}a^{4}-6a^{\left(  4\right)
}\ddot{a}a^{4}+12a^{\left(  4\right)  }\dot{a}^{2}a^{3}\nonumber\\
&  +3~\dddot{a}^{2}a^{4}+18\dddot{a}\ddot{a}\dot{a}a^{3}-78\dddot{a}\dot
{a}^{3}a^{2}\nonumber\\
&  -6\ddot{a}^{3}a^{3}-39\ddot{a}^{2}\dot{a}^{2}a^{2}+78\ddot{a}\dot{a}%
^{4}a-180\dot{a}^{6} \label{F1(a)}%
\end{align}
In order to solve Eq. (\ref{Fried a B}), we define the non-dimensional
variables%
\begin{align}
\mathcal{H}  &  \equiv\frac{\dot{a}}{H_{0}}=a\frac{H}{H_{0}}~;\label{x}\\
Q  &  \equiv\frac{\mathcal{\dot{H}}}{H_{0}}=-a\frac{H^{2}}{H_{0}^{2}%
}q~;\label{Q}\\
J  &  \equiv\frac{\dot{Q}}{H_{0}}=a\frac{H^{3}}{H_{0}^{3}}j~;\label{J}\\
S  &  \equiv\frac{\dot{J}}{H_{0}}=a\frac{H^{4}}{H_{0}^{4}}s~, \label{S}%
\end{align}
which are related to the Hubble function, deceleration parameter, jerk and
snap. In fact, when these variables are calculated at the present time,
$t=t_{0}$, they give precisely the well known cosmological parameters:
$\mathcal{H}\left(  t_{0}\right)  =\mathcal{H}_{0}=a_{0}=1$, $Q_{0}=-q_{0}$,
$J_{0}=j_{0}$\ and $S_{0}=s_{0}$. In the set of variables (\ref{x})-(\ref{S}),
one variable is essentially the derivative of the previous one, therefore they
can be used to turn the fifth-order differential equation Eq.(\ref{Fried a B})
into five first order differential equations to be solved simultaneously.
Following the Method of Characteristics, we choose the scale factor $a$\ as
the independent variable of the model (instead of the cosmic time $t$),
obtaining the system%
\begin{equation}
\left\{
\begin{array}
[c]{l}%
\frac{d\mathcal{H}}{dz}=-\frac{1}{\left(  1+z\right)  ^{2}}\frac
{Q}{\mathcal{H}}\\
\frac{dQ}{dz}=-\frac{1}{\left(  1+z\right)  ^{2}}\frac{J}{\mathcal{H}}\\
\frac{dJ}{dz}=-\frac{1}{\left(  1+z\right)  ^{2}}\frac{S}{\mathcal{H}}\\
\frac{dS}{dz}=\frac{-1}{\left(  1+z\right)  ^{2}}F_{3}\left(  z;\mathcal{H}%
,Q,J,S\right)
\end{array}
\right.  ~, \label{system(z)}%
\end{equation}
with%
\begin{align}
F_{3}\left(  z;\mathcal{H},Q,J,S\right)   &  =-2S\left(  1+z\right)
+\frac{SQ}{\mathcal{H}^{2}}\nonumber\\
&  -\frac{1}{2}\left(  \frac{J}{\mathcal{H}}\right)  ^{2}-3\frac{JQ\left(
1+z\right)  }{\mathcal{H}}+13J\mathcal{H}\left(  1+z\right)  ^{2}\nonumber\\
&  +\frac{Q^{3}\left(  1+z\right)  }{\mathcal{H}^{2}}+\frac{13}{2}Q^{2}\left(
1+z\right)  ^{2}-13Q\mathcal{H}^{2}\left(  1+z\right)  ^{3}\nonumber\\
&  +30\mathcal{H}^{4}\left(  1+z\right)  ^{4}+\frac{\Omega_{m0}}{6B}%
\frac{\left(  1+z\right)  }{\mathcal{H}^{2}}-\frac{1}{6B}~, \label{F3}%
\end{align}
where we have used the relationship between the redshift and the scale factor,%
\begin{equation}
\left(  1+z\right)  =\frac{1}{a}~. \label{redshift}%
\end{equation}

The system of Eq.(\ref{system(z)}) is solved numerically, which is indeed a
far more simpler task than dealing directly with Eq. (\ref{Fried a B}). This
leads to a function $\mathcal{H}\left(  z\right)  $ (essentially equal to
$\dot{a}$) for each set $[\mathcal{M}]=\{q_{0},j_{0},s_{0},B,\Omega_{m0}%
\}$\ of cosmological parameters. This function is used to obtain an
interpolated function for the luminosity distance $d_{L}=d_{L}\left(
z,[\mathcal{M}]\right)  $\ of our model -- defined bellow -- which is to be
fitted to the data from SN Ia observations and from $f$-gas measurements.

\section{Fitting the model to the SN Ia and $f$-gas data \label{sec-DataFit}}

\subsection{SN Ia data}

In order to relate the proposed model to the observational data, one has to
build the luminosity distance $d_{L}$ \cite{Amendola2010},%
\begin{equation}
d_{L}=\left(  1+z\right)
{\displaystyle\int\limits_{0}^{z}}
\frac{d\bar{z}}{H\left(  \bar{z}\right)  }=\frac{\left(  1+z\right)  }{H_{0}}%
{\displaystyle\int\limits_{0}^{z}}
\frac{d\bar{z}}{\left(  1+\bar{z}\right)  ~\mathcal{H}\left(  \bar{z}\right)
}~. \label{dL(z)}%
\end{equation}
Our task is to solve the system Eq.(\ref{system(z)}) for obtaining
$\mathcal{H}$, then to integrate numerically the non-dimensional luminosity
distance,%
\begin{equation}
d_{h}\left(  z;[\mathcal{M}]\right)  =\frac{H_{0}}{c}d_{L}\left(
z;[\mathcal{M}]\right)  ~,\;\;\;\;\;\;\left(  c=1\right)  \label{dh(z)}%
\end{equation}
for each set of cosmological parameters $[\mathcal{M}]=\{q_{0},j_{0}%
,s_{0},B,\Omega_{m0}\}$.

Function $d_{h}$\ appears in the definition of the distance modulus,%
\begin{equation}
\mu_{r}=5\log_{10}d_{h}+5\log_{10}\left(  \frac{c}{H_{0}d_{0}}\right)  ~,
\label{mu_r}%
\end{equation}
where $d_{0}$\ is the reference distance of $10$ pc. Observational data give
$\mu_{r}$ of the supernovae and their corresponding redshift\ $z$.

According to the likelihood method the probability distribution of the data%
\begin{equation}%
\mathcal{L}%
_{SNIa}\left(  \left[  \mathcal{M}\right]  |\left\{  z_{i},\mu_{r_{i}%
}\right\}  \right)  =N_{SNIa}e^{-\frac{\chi_{SNIa}^{2}\left(  \mathcal{[M]}%
\right)  }{2}} \label{L SNIa}%
\end{equation}
($N_{SNIa}$ is a normalization constant) is the one that minimizes
\begin{equation}
\chi_{SNIa}^{2}\left(  [\mathcal{M}]\right)  =%
{\displaystyle\sum\limits_{i}}
\frac{\left(  \mu_{r_{i}}-5\log d_{h}\left(  z_{i},[\mathcal{M}]\right)
-5\log_{10}\left(  \frac{c}{H_{0}d_{0}}\right)  \right)  ^{2}}{\sigma_{\mu
_{i}}^{2}+\sigma_{ext}^{2}}~, \label{Chi2 SNIa}%
\end{equation}
where $\sigma_{\mu_{i}}$\ is the uncertainty related to the distance modulus
$\mu_{r_{i}}$ of the $i$-th supernova. The data set used in our calculations
is composed with the 557 supernovae of the Union 2 compilation \cite{Union2}.
Uncertainty $\sigma_{ext}$\ is given by%
\begin{equation}
\sigma_{ext}^{2}=\sigma_{\mu_{v}}^{2}+\sigma_{lens}^{2}~, \label{sigma ext}%
\end{equation}
i.e., it is the combination of the error $\sigma_{v}$ in the magnitude due to
host galaxy peculiar velocities (of order of $300%
\operatorname{km}%
\operatorname{s}%
^{-1}$),%
\begin{equation}
\sigma_{\mu_{v}}=\frac{2.172}{z}\frac{\sigma_{v}}{c}\text{\ \ \ with\ \ \ }%
\sigma_{v}=300%
\operatorname{km}%
\operatorname{s}%
^{-1}~, \label{sigma mu v}%
\end{equation}
and the uncertainty from gravitational lensing
\begin{equation}
\sigma_{lens}=0.093z \label{sigma lens}%
\end{equation}
which is progressively important at increasing SN redshifts. Further detail
can be found in Refs. \cite{Medeiros2012,Holz2005}, for example.

\subsection{Galaxy cluster X-ray data}

Measurements of the fraction of the mass of gas in galaxy clusters in the
X-ray band of the spectrum constrain efficiently the ratio $\frac{\Omega_{b0}%
}{\Omega_{m0}}$ and exhibit a lower dependence on the values of other
cosmological parameters \cite{Schindler2002,Allen2008}.\footnote{$\Omega
_{b0}=\rho_{b}/\rho_{c}$ is the non-dimensional baryon density parameter.}
According to Ref.\cite{Allen2008}, the mass fraction of gas $f_{gas}$ relates
to $\frac{\Omega_{b0}}{\Omega_{m0}}$\ as%
\begin{equation}
f_{gas}^{\Lambda CDM}\left(  z\right)  =\frac{KA\gamma b\left(  z\right)
}{1+s\left(  z\right)  }\left(  \frac{\Omega_{b0}}{\Omega_{m0}}\right)
\left[  \frac{d_{A}^{\Lambda CDM}\left(  z\right)  }{d_{A}\left(  z\right)
}\right]  ^{1,5}~, \label{f_gas}%
\end{equation}
where $d_{A}$ is the angular distance for $\kappa=0$,%
\begin{equation}
d_{A}\left(  z\right)  =\frac{d_{L}\left(  z\right)  }{\left(  1+z\right)
^{2}}=\frac{1}{\left(  1+z\right)  }\frac{1}{H_{0}}%
{\displaystyle\int\limits_{0}^{z}}
\frac{d\bar{z}}{\left(  1+\bar{z}\right)  \mathcal{H}\left(  \bar
{z};[\mathcal{M}]\right)  }~. \label{dA}%
\end{equation}
The term $d_{A}^{\Lambda CDM}\left(  z\right)  $ is the angular distance for
the standard $\Lambda$CDM model, for which $\Omega_{m0}=0.3$, $\Omega
_{\Lambda}=0.7$ and $H_{0}^{70}=70%
\operatorname{km}%
\operatorname{s}%
^{-1}$Mpc$^{-1}$, i.e.%
\begin{equation}
d_{A}^{\Lambda CDM}\left(  z\right)  =\frac{1}{\left(  1+z\right)  }\frac
{1}{H_{0}}%
{\displaystyle\int\limits_{0}^{z}}
\frac{d\bar{z}}{\left[  \Omega_{m0}\left(  1+\bar{z}\right)  ^{3}%
+\Omega_{\Lambda}\right]  ^{1/2}}. \label{dA LCDM}%
\end{equation}
The other factors in Eq. (\ref{f_gas}) are defined bellow:

\begin{itemize}
\item $K$ is a constant due to the calibration of the instruments observing in
X-ray. Following Ref. \cite{Allen2008}, we take $K=1.0\pm0.1$.

\item Constant $A$ measures the angular correction on the path of an light ray
propagating in the cluster due to the difference between $\Lambda$CDM model
and the model that we want to fit to the data. Ref. \cite{Allen2008} argues
that for relatively low red-shifts (as those involved here) this factor can be
neglected, i.e. taken as $A=1$.

\item $\gamma$ accounts for the influence of the non-thermal pressure within
the clusters. According to\ Ref.\cite{Allen2008}, one may set $1.0<\gamma
<1.1$. Our choice is $\gamma=1.050\pm0.029$.

\item $b\left(  z\right)  $ is the bias factor related to the thermodynamical
history of the inter-cluster gas. Once again due to arguments given in
Ref.\cite{Allen2008}, this factor can be modeled as\ $b\left(  z\right)
=b_{0}\left(  1+\alpha_{b}z\right)  $ where $b_{0}=0.83\pm0.04$ and
$-0.1<\alpha_{b}<0.1$. We set $\alpha_{b}=0.000\pm0.058$.

\item Parameter $s\left(  z\right)  =s_{0}\left(  1+\alpha_{s}z\right)  $
describes the fraction of baryonic mass in the stars.\footnote{Notice that
here $s_{0}$ is not the snap, defined in Eq. (\ref{snap}). Even at risk of
confusion, we decided to maintain the notation used in Ref. \cite{Allen2008}.}
Ref. \cite{Allen2008} uses $s_{0}=\left(  0.16\pm0.05\right)  h_{70}^{1/2}$
and $-0.2<\alpha_{s}<0.2$. Here
\begin{equation}
h_{70}=\frac{H_{0}}{H_{0}^{70}\text{ }}~, \label{h70}%
\end{equation}
and we will set $\alpha_{s}=0.000\pm0.115$. Analogously to Eq.(\ref{h70}), we
define%
\begin{equation}
h=\frac{H_{0}}{\left(  100%
\operatorname{km}%
\operatorname{s}%
^{-1}Mpc^{-1}\right)  }~, \label{h}%
\end{equation}
that will be required bellow.

\item The baryon parameter $\Omega_{b0}$ is set to%
\begin{equation}
\Omega_{b0}h^{2}=0.0214\pm0.002 \label{Omega_b0}%
\end{equation}
following the constraints imposed by the primordial nucleosynthesis on the
ratio $D/H$ \cite{Kirkman2003}.
\end{itemize}

The uncertainties in the set of parameters $\{b_{0},s_{0},K,\Omega_{b0}%
h^{2},\alpha_{s},\alpha_{b},\gamma\}$ are propagated to build the uncertainty
of $f_{gas}^{\Lambda CDM}$. It is worth mentioning that the quantities
$\left\{  b_{0},s_{0},K,\Omega_{b0}h^{2}\right\}  $\ have Gaussian
uncertainties. On the other hand, the parameters $\left\{  \alpha_{s}%
,\alpha_{b},\gamma\right\}  $\ present a rectangular distribution, e.g.
$\gamma$\ has equal probability of assuming values in the interval $\left[
1.0,1.1\right]  $. The uncertainty of the parameters obeying the rectangular
distribution is estimated as the half of their variation interval divided by
$\sqrt{3}$. Maybe it is useful to remark that in the case of $s_{0}$, we
adopted $h_{70}^{1/2}\simeq1$. Moreover, $\frac{h^{2}}{h_{70}^{2}}=\left(
\frac{0.7}{1}\right)  ^{2}\cong0.49$.

Ref. \cite{Allen2008} provides data for the fraction of gas $f_{gas}$ and its
uncertainty $\sigma_{f}$ of 42 galaxy clusters with redshift $z$. These data
can be used to determine the free parameter of our model by minimizing
\begin{equation}
\chi_{gas}^{2}\left(  [\mathcal{M}]\right)  =%
{\displaystyle\sum\limits_{i}}
\frac{\left(  f_{gas}^{i}-f_{gas}^{\Lambda CDM}\left(  z,[\mathcal{M}]\right)
\right)  ^{2}}{\sigma_{f_{i}}^{2}} \label{Chi2 f-gas}%
\end{equation}
Notice that the dependence of $\chi^{2}$ on the set $\left[  \mathcal{M}%
\right]  $ of cosmological parameters occurs through the function
$f_{gas}^{\Lambda CDM}$ given by Eq.(\ref{f_gas}), i.e. ultimately through
$d_{A}\left(  z,\left[  \mathcal{M}\right]  \right)  $ -- Eq. \ref{dA}.

Function $\chi_{gas}^{2}$\ is related to the distribution
\begin{equation}%
\mathcal{L}%
_{gas}\left(  \left[  \mathcal{M}\right]  \right)  =N_{gas}e^{-\frac
{\chi_{gas}^{2}\left(  \mathcal{[M]}\right)  }{2}} \label{L gas}%
\end{equation}
which should be maximized.

\subsection{Combining SN Ia and $f$-gas data sets}

The majority of papers that present cosmological models tested through SN Ia
data fit -- see e.g. Refs. \cite{Gabi2007,Medeiros2012}\ -- perform an
analitical marginalization with respect to $H_{0}$. On the other hand,
constraint analyses using $f$-gas data adopt a fixed value for $H_{0}$. Our
basic reference on $f$-gas treatment, namely Ref.\cite{Allen2008}, choose
$H_{0}=70%
\operatorname{km}%
\operatorname{s}%
^{-1}$Mpc$^{-1}$. As we will consider SN Ia and $f$-gas data sets in the same
footing, it is convenient to adopt $H_{0}=70%
\operatorname{km}%
\operatorname{s}%
^{-1}$Mpc$^{-1}$ as a fixed quantity in our calculations (and not the most
up-to-date value of $\left(  73.8\pm2.4\right)
\operatorname{km}%
\operatorname{s}%
^{-1}$Mpc$^{-1}$\ found in \cite{Riess2011}).

We want to maximize the probability distributions given in Eqs.(\ref{L SNIa})
and (\ref{L gas}) simultaneously, which is done by maximixing%
\begin{equation}%
\mathcal{L}%
\left(  \left[  \mathcal{M}\right]  \right)  =%
\mathcal{L}%
_{SNIa}\left(  \left[  \mathcal{M}\right]  \right)
\mathcal{L}%
_{gas}\left(  \left[  \mathcal{M}\right]  \right)  ~, \label{L}%
\end{equation}
or minimizing%
\begin{equation}
\chi^{2}\left(  [\mathcal{M}]\right)  =\chi_{SNIa}^{2}\left(  [\mathcal{M}%
]\right)  +\chi_{gas}^{2}\left(  [\mathcal{M}]\right)  ~. \label{Chi2}%
\end{equation}

\section{Results \label{sec-Results}}

\subsection{Estimation of the free parameters of the $f\left(  R,\nabla
R\right)  $-model \label{sec-ParametersFit}}

\begin{figure}[h]
\begin{center}
\includegraphics[width=15cm]{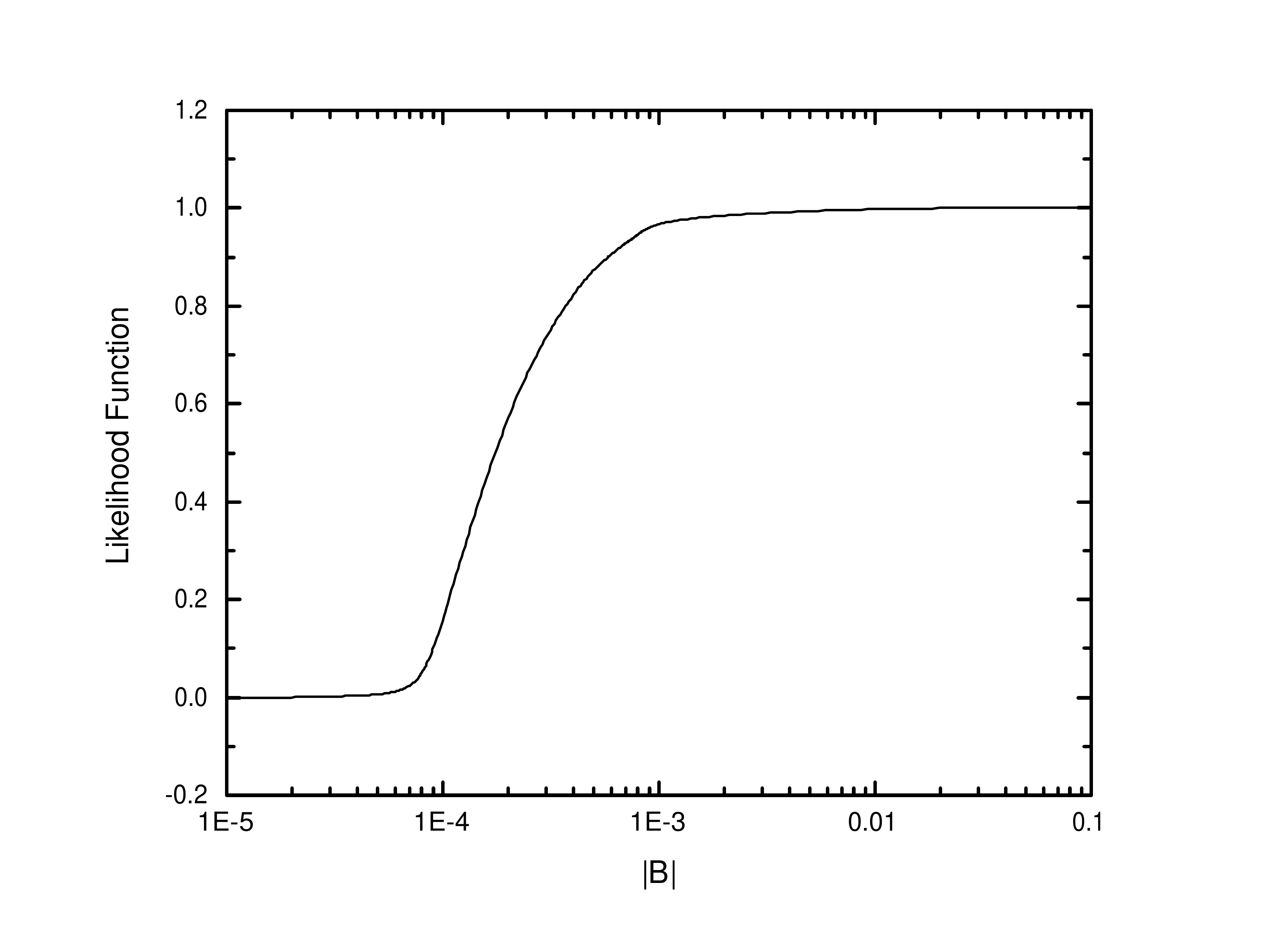}
\end{center}
\caption{Sketch\ of the likelihood function $\mathcal{L} \left(  \left[
\mathcal{M}\right]  \right) $ as a function of the absolute value of the
coupling parameter $B$. The other parameters $\left\{  q_{0},j_{0}%
,s_{0},\Omega_{m0}\right\} $ were fixed in their best-fit values.}%
\label{FigB}%
\end{figure}

The parameters to be ajusted are of two types. The kinematic parameters are
$\left\{  q_{0},j_{0},s_{0}\right\}  $;\footnote{We emphasize that $\left\{
q_{0},j_{0},s_{0}\right\} $ are precisely the values of $\left\{
-Q,J,S\right\}  $ calculated at the present time $t=t_{0}$.} the cosmological
parameters are $\left\{  \Omega_{m0},B\right\} $. Maximization of the
likelihood function $\mathcal{L} \left(  \left[  \mathcal{M}\right]  \right) $
-- Eq. (\ref{L})\ -- determines all kinematic parameters and one cosmological
parameter, namely $\left\{  q_{0},j_{0},s_{0},\Omega_{m0}\right\}  $.
Parameter $B$ is not fixed by statistical treatment of the data, and the
reason for that is evident in the plot of Fig. \ref{FigB}.

The curve of $%
\mathcal{L}%
\left(  \left\vert B\right\vert \right)  $ is not Gaussian but rather shows a
flat region at high values of $\left\vert B\right\vert $. Indeed, values of
$\left\vert B\right\vert $ greater than $10^{-3}$\ are almost equally probable
and more favorable than the $\left\vert B\right\vert <10^{-3}$. This shows it
is not possible to determine $B$ by maximizing its likelihood function.
Instead, we choose four values of $B$ to proceed the analysis leading to the
determination of the free parameters of the model. These values are $B=\left(
\pm10^{-3},\pm10^{-1}\right)  $, so that we have two values of $B$ in the
beginning of the plateau ($\left\vert B\right\vert =10^{-3}$)\ and two of them
far from the beginning of the plateau ($\left\vert B\right\vert =10^{-1}$).
These choices are only partially arbitrary: we have done preliminary
calculations with other values of $\left\vert B\right\vert $\ ($\left\vert
B\right\vert =1,10,100$) and the results for the best-fit calculations of
$\left\{  q_{0},j_{0},s_{0},\Omega_{m0}\right\}  $ are almost the same as
those obtained by using $\left\vert B\right\vert =10^{-1}$.

The standard statistical analysis leads to the best-fit values of the
parameters shown in Table \ref{table-Parameters}.

\begin{table}[pth]
\caption{Fit results of the parameters of our $f\left(  R,\nabla R\right)
$-model. The uncertainties represent a 68\% confidence level (CL).}%
\label{table-Parameters}
\begin{center}%
\begin{tabular}
[c]{||c||c||c||c||c||}\hline\hline
$B$ & $q_{0}$ & $j_{0}$ & $s_{0}$ & $\Omega_{m0}$\\\hline\hline
$-10^{-1}$ & $-0.51_{-0.14}^{+0.14}$ & $0.3_{-2.5}^{+1.7}$ & $0.4_{-11.1}%
^{+9.3}$ & $0.301_{-0.010}^{+0.010}$\\\hline\hline
$-10^{-3}$ & $-0.55_{-0.14}^{+0.14}$ & $1.4_{-2.2}^{+2.1}$ & $18.5_{-11.0}%
^{+9.8}$ & $0.301_{-0.010}^{+0.011}$\\\hline\hline
$+10^{-3}$ & $-0.46_{-0.14}^{+0.14}$ & $-1.6_{-2.1}^{+2.1}$ & $-19.4_{-10.3}%
^{+9.7}$ & $0.301_{-0.010}^{+0.010}$\\\hline\hline
$+10^{-1}$ & $-0.51_{-0.14}^{+0.14}$ & $0.3_{-2.5}^{+1.7}$ & $0.4_{-11.0}%
^{+9.3}$ & $0.301_{-0.010}^{+0.010}$\\\hline\hline
\end{tabular}
\end{center}
\end{table}


It is clear from Table \ref{table-Parameters} that our higher order
cosmological model presents an acceleration at present time: the deceleration
parameter $q_{0}$ is negative for all the values of the coupling constant $B$.
This is a robust result indeed once we have got negative values for $q_{0}$
with $99.7\%$ confidence level.

Both SN Ia and $f$-gas data sets give information about astrophysical objects
at relatively low redshifts ($z\lesssim1.4$). That is the reason for the high
values of the uncertainties on the parameter $j_{0}$\ and $s_{0}$. Positive
values of $j_{0}$\ mean that the tendency is the increasing of the rate of
cosmic acceleration. Because the jerk parameter $j_{0}$\ does not exhibit the
same sign for all values of $B$, it is not possible to argue in favor of an
overall behavior for the cosmic acceleration today. This statement is also
applicable to the snap $s_{0}$ obtained with $B=\pm10^{-1}$. However, the
uncertainties in the value of $s_{0}$ for $B=-10^{-3}$\ do indicate that it is
positive within 68\% CL. The negative value of $s_{0}$\ for $B=+10^{-3}$\ is
also guaranteed within the same CL. This difference in the sign of $s_{0}%
$\ for $B=-10^{-3}$ and $B=+10^{-3}$\ leads to distinct dynamics for the scale
factor towards the future, as we shall discuss in Section \ref{sec-Behaviour}.

The best-fits for $\Omega_{m0}$\ are statistically the same for all the chosen
values of $B$. This is linked to the fact that the $f$-gas data produce a
sharp determination of this parameter. Notice that the value $\Omega_{m0}$\ of
our model is compatible with that of $\Lambda$CDM model as reported in
Ref.\cite{Union2}\ within 95\% CL. Inspection of Table \ref{table-Parameters}
indicates that the Gaussian-like distributions centered in each parameter of
the set $\left\{  q_{0},j_{0},s_{0},\Omega_{m0}\right\}  $ for $B=-10^{-1}%
$\ overlap the distributions centered in $\left\{  q_{0},j_{0},s_{0}%
,\Omega_{m0}\right\}  $ for $B=+10^{-1}$; this overlap of distribution curves
occurs within 68\% CL of each distribution. Moreover, this overlap of
distributions is greater than the one observed in the case of the sets
$\left\{  q_{0},j_{0},s_{0},\Omega_{m0}\right\}  $ for $B=-10^{-3}$\ and
$B=+10^{-3}$, which occurs within 95\% CL. Hence, the greater the value of
$\left\vert B\right\vert $, the greater the overlap of distribution and the
less important is our choice of the value of $B$. This emphasizes what was
said before about the choice of values for $B$.

The confidence regions (CR) in the $\left(  q_{0},j_{0}\right)  $ and $\left(
q_{0},\Omega_{m0}\right)  $ planes for each one of the four values of $B$ are
shown in Figs. \ref{RegBminus01}-\ref{Fig-RegionsBplus01}.

\begin{center}

\begin{figure}[ptbh]
\includegraphics[width=8cm]{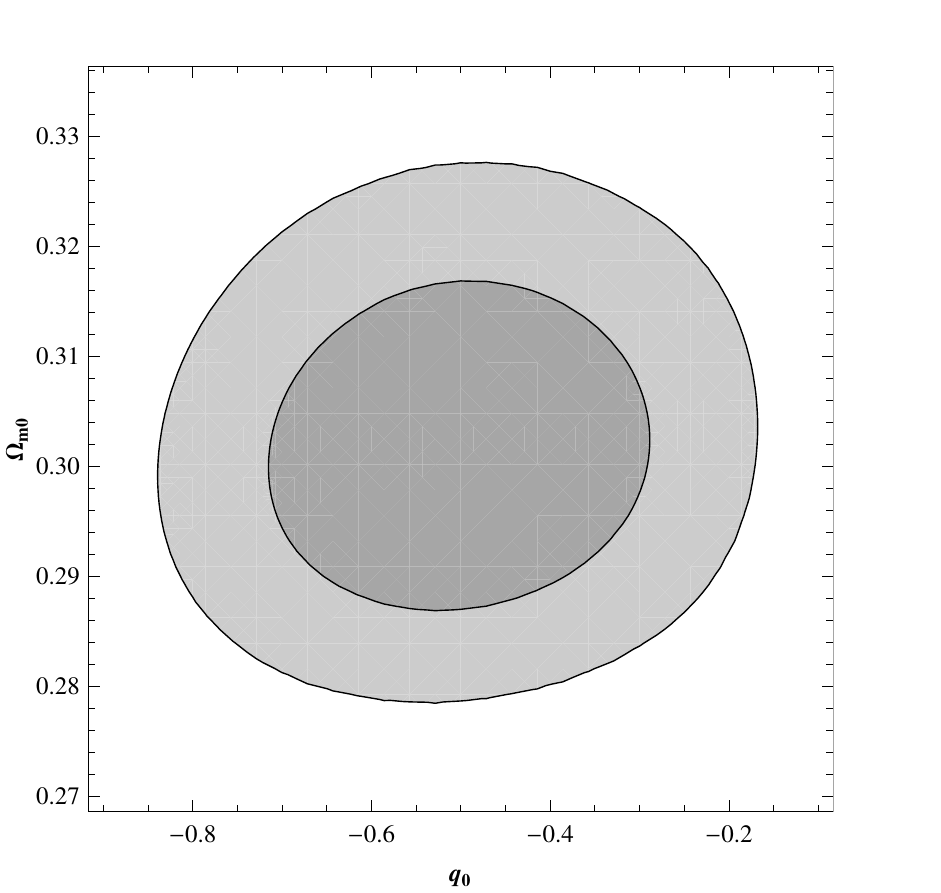}
\includegraphics[width=8cm]{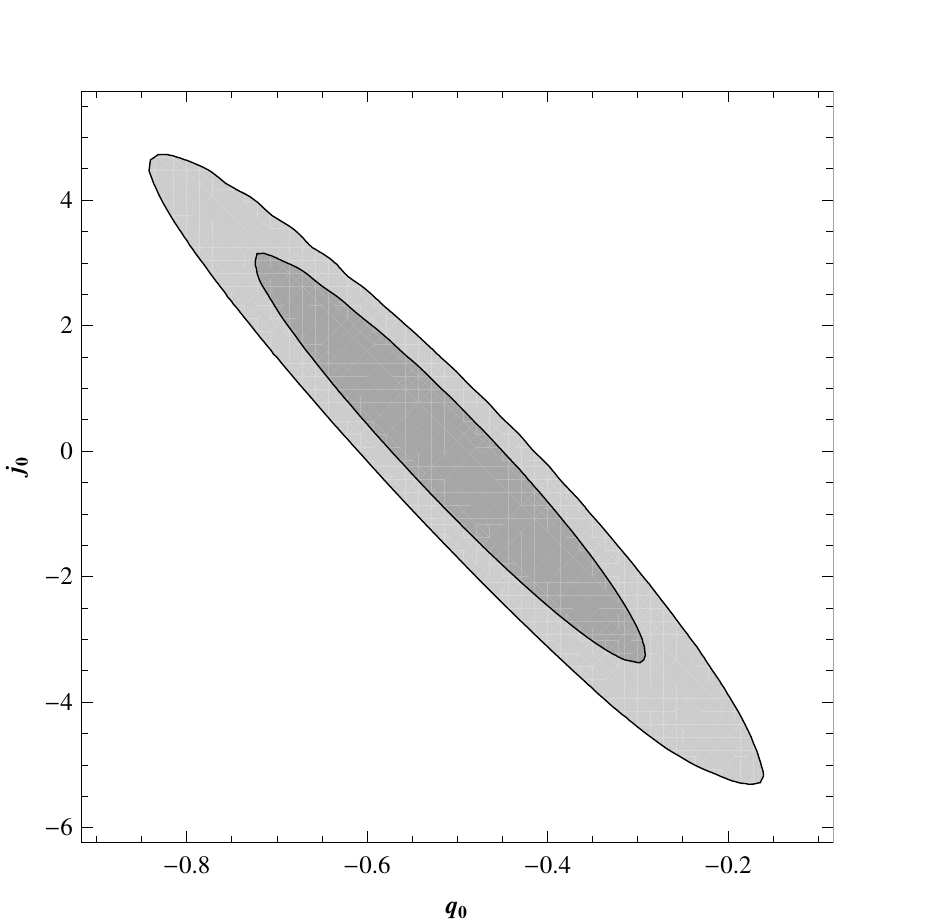}\caption{Confidence regions in
the $\left(  q_{0},\Omega_{m0}\right)  $ and $\left(  q_{0},j_{0}\right)  $
planes for $B=-10^{-1}$ from SN Ia and $f$-gas data sets. The inner contour
contains the 68\% CR, whilst the outer curves confines the 95\% CR.}%
\label{RegBminus01}%
\end{figure}

\begin{figure}[ptbh]
\includegraphics[width=8cm]{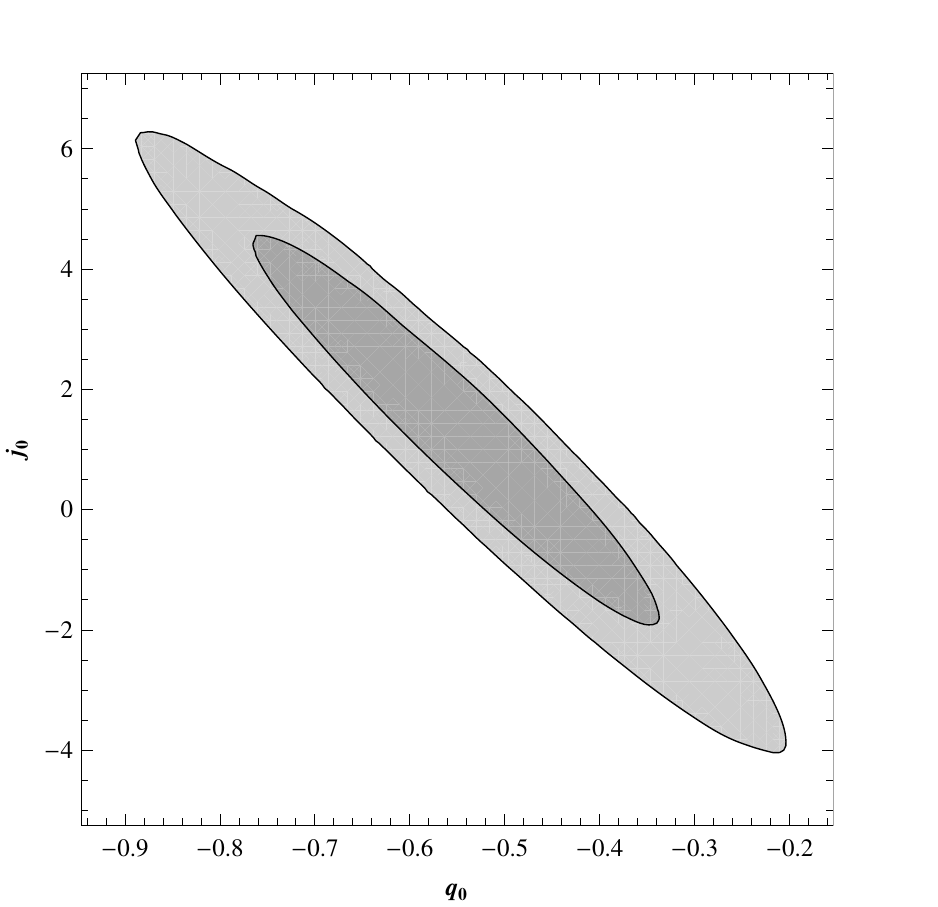}
\includegraphics[width=8cm]{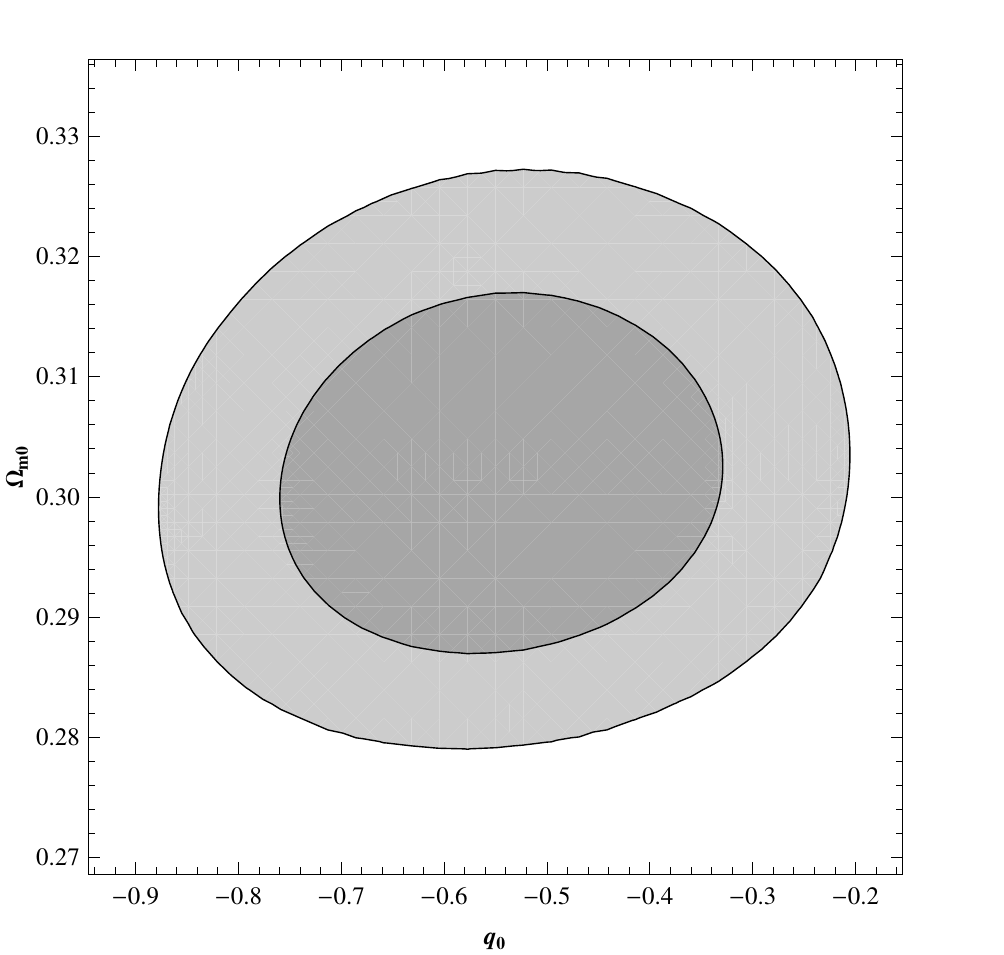}\caption{68\% and 95\%
confidence regions in the $\left(  q_{0},j_{0}\right)  $ and $\left(
q_{0},\Omega_{m0}\right)  $ planes for $B=-10^{-3}$.}%
\label{RegBminusMinus3}%
\end{figure}

\begin{figure}[ptbh]
\includegraphics[width=8cm]{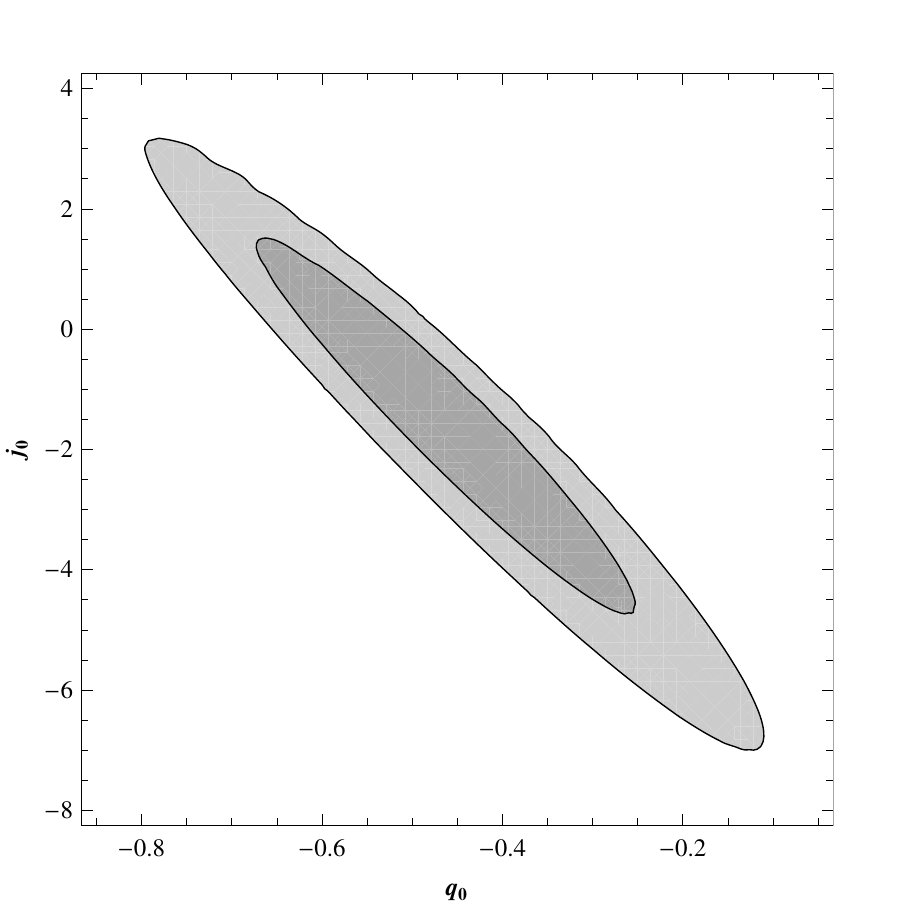}
\includegraphics[width=8cm]{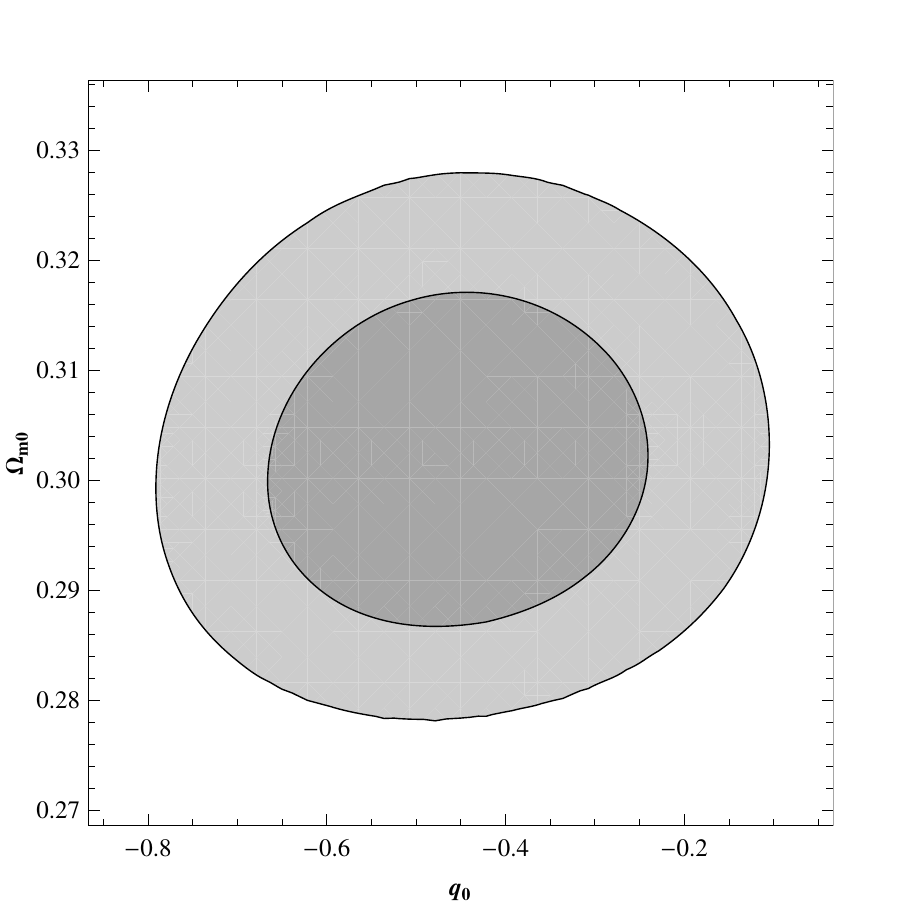}\caption{Confidence regions in
the plots $\left(  q_{0},j_{0}\right)  $ and $\left(  q_{0},\Omega
_{m0}\right)  $ planes for the value $B=+10^{-3}$ of the coupling constant.}%
\label{RegBplus0001}%
\end{figure}

\begin{figure}[ptbh]
\includegraphics[width=8cm]{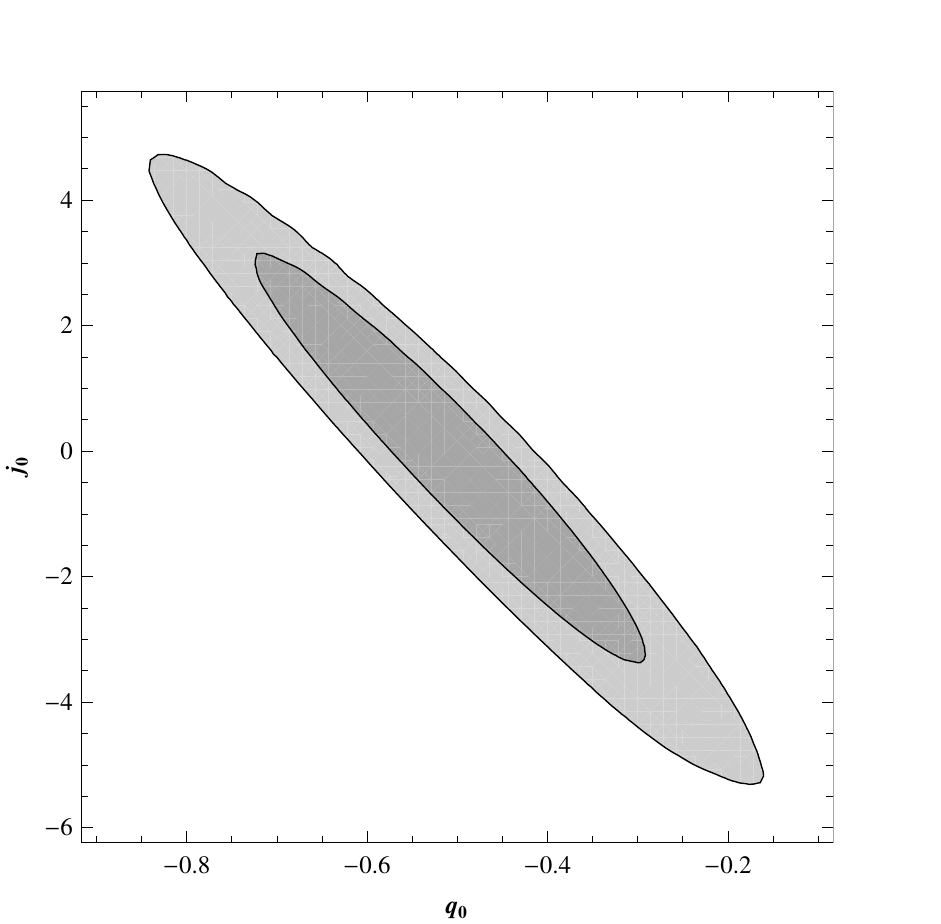}
\includegraphics[width=8cm]{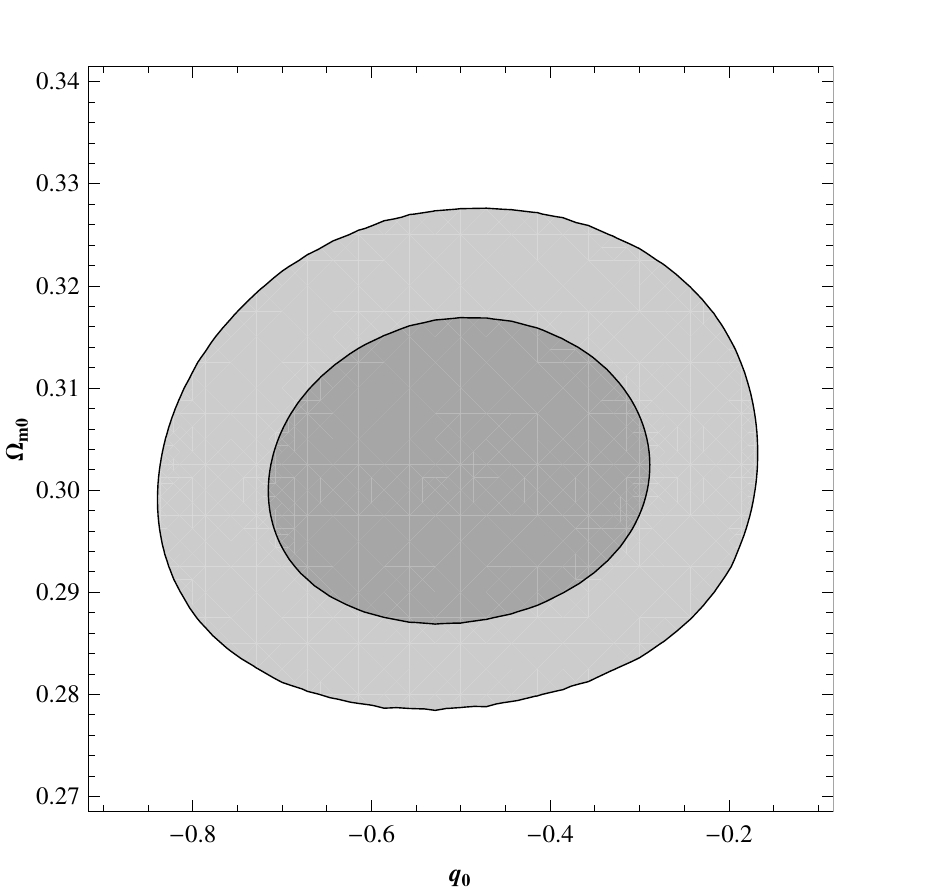}\caption{CR in the planes
$\left(  q_{0},j_{0}\right)  $ and $\left(  q_{0},\Omega_{m0}\right)  $ for
$B=+10^{-1}$.}%
\label{Fig-RegionsBplus01}%
\end{figure}

\end{center}

The shape of the confidence regions for $\left(  q_{0},j_{0}\right)
$\ indicates that the deceleration parameter $q_{0}$\ and the jerk $j_{0}%
$\ are correlated parameters; this occurs for all the values of $B$. The four
plots $\left(  q_{0},j_{0}\right)  $ point to an accelerated universe at
recent times within 95\% CL, but the change in the acceleration rate takes
positive and negative values with equal probability.

Conversely, the shape of the plots of $\left(  q_{0},\Omega_{m0}\right)  $\ in
Figs. \ref{RegBminus01}--\ref{Fig-RegionsBplus01} indicate that $q_{0}$\ and
$\Omega_{m0}$\ are weakly correlated parameters. Notice that the difference
between the maximum and the minimum values of $\Omega_{m0}$\ in the range of
the plots $q_{0}\times\Omega_{m0}$ is only 0.05 with 95\% CL: this confirms
that the parameter $\Omega_{m0}$\ is sharply determined by the best-fit
procedure (something that is consistent with the fact that we have employed
$f$-gas data to build the likelihood function).

In order to quantify the correlation between the parameters $\left(
q_{0},\,j_{0}\right)  $ and $\left(  q_{0},\,\Omega_{m0}\right)  $, the
covariances $\sigma_{q_{0}j_{0}}$, $\sigma_{q_{0}\Omega_{m0}}$ and correlation
coefficients $r_{q_{0}j_{0}}$, $r_{q_{0}\Omega_{m0}}$ were evaluated according
to Ref.{ \cite{Beringer2012}}. The mean values and their respective standard
deviations used to calculate the covariances and correlation coefficients were
those given in Table \ref{table-Covariances}. The same table shows the values
obtained for the covariances and correlation functions for each value of $B$
taken in this paper.

\begin{table}[pth]
\caption{Values of mean and standard deviation of the parameters $\left\{
q_{0},j_{0},\Omega_{m0},s_{0}\right\}  $ for different values of $B$. The
values of $s_{0}$, although irrelevant for the evaluation of the parameters of
interest, were included for completeness. The last four columns show the
values of covariances $\sigma_{q_{0}j_{0}}$, $\sigma_{q_{0}\Omega_{m0}}$ and
correlation coefficients $r_{q_{0}j_{0}}$, $r_{q_{0}\Omega_{m0}}$.}%
\label{table-Covariances}
\begin{center}%
\begin{tabular}
[c]{||c||c||c||c||c||c||c||c||c||}\hline\hline
$B$ & $q_{0}$ & $j_{0}$ & $\Omega_{m0}$ & $s_{0}$ & $\sigma_{q_{0}j_{0}}$ &
$r_{q_{0}j_{0}}$ & $\sigma_{q_{0}\Omega_{m0}}$ & $r_{q_{0}\Omega_{m0}}%
$\\\hline\hline
$-10^{-1}$ & $-0.50\pm0.14$ & $-0.2\pm2.1$ & $0.302\pm0.010$ & $-1.5\pm9.8$ &
$-0.269$ & $-0.954$ & $0.00011$ & $0.0817$\\\hline\hline
$-10^{-3}$ & $-0.54\pm0.14$ & $1.2\pm2.1$ & $0.299\pm0.010$ & $17\pm10$ &
$-0.267$ & $-0.952$ & $0.00012$ & $0.0812$\\\hline\hline
$+10^{-3}$ & $-0.45\pm0.14$ & $-1.7\pm2.1$ & $0.302\pm0.010$ & $-20.9\pm9.9$ &
$-0.272$ & $-0.956$ & $0.00011$ & $0.0753$\\\hline\hline
$+10^{-1}$ & $0.50\pm0.14$ & $-0.2\pm2.1$ & $0.302\pm0.010$ & $-1.5\pm9.9$ &
$-0.258$ & $-0.954$ & $0.00011$ & $0.0817$\\\hline\hline
\end{tabular}
\end{center}
\end{table}


The values of the correlation coefficient presented in Table
\ref{table-Covariances} confirm what was stated before: The parameters $q_{0}$
and $j_{0}$ are strongly (negatively) correlated while $q_{0}$ and
$\Omega_{m0}$ are weakly correlated for all values of B considered here.

Comparison of the results presented in Table \ref{table-Covariances} and Table
\ref{table-Parameters} shows that there is no statistical difference between
considering the mode and a confidence interval of 68\% (Table
\ref{table-Parameters}) from taking the mean and standard deviation. This
emphasizes the quasi-gaussian behaviour of the probability function.


\subsection{Behaviour of the $f\left(  R,\nabla R\right)  $-model with time
\label{sec-Behaviour}}

We can investigate the evolution of our higher order model in time by plotting
the function $\mathcal{H}=\dot{a}/H_{0}$ as a function of $z$. This is done
setting the parameters at their best-fit results for each value of $B$.

\begin{figure}[ptbh]
\begin{center}
\includegraphics[
natheight=4.715800in,
natwidth=6.000100in,
height=3.0727in,
width=3.9038in
]
{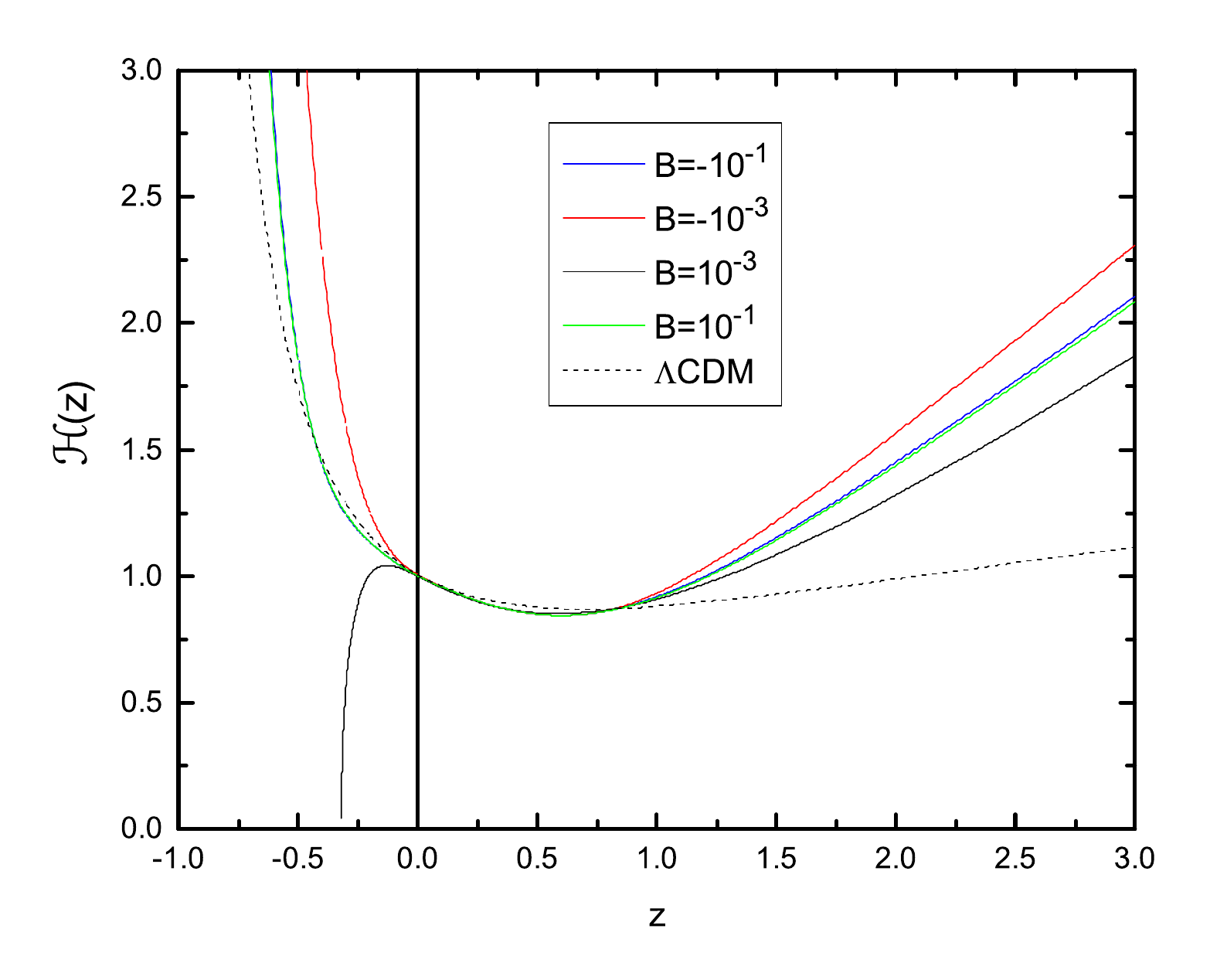}
\end{center}
\caption{Plot of the function $\mathcal{H}\left(  z\right)  $ accounting for
the \emph{best-fit} parameters of the higher order model. This is essentially
the curve of $\dot{a}$\ as a function of $z$\ for our model -- colored lines
-- and for the standard $\Lambda$CDM\ model -- dashed line. The notation is:
$B=-10^{-1}$ is the blue curve; $B=-10^{-3}$, red line; $B=+10^{-3}$, black
curve; and $B=+10^{-1}$, green line.}%
\label{Fig-Behaviour}%
\end{figure}


There is agreement between our model -- regardless the value of $B$ -- and the
standard $\Lambda$CDM\ model in the region $0\lesssim z\lesssim1$. In
particular, the curves of $\mathcal{H}\left(  z\right)  $ for $B=-10^{-1}%
$\ and $B=+10^{-1}$\ almost coincide: the difference between them is only
noticeable at $z\simeq3$. The transition from a decelerated phase to an
accelerated expansion in the higher order model occurs at $z\simeq0.6$ for all
values of $B$.

Towards the future ($t>t_{0}$) the redshift takes negative values, $-1<z<0$.
In this region our model present different dynamics depending on the value of
the coupling constant. For $B=-10^{-1}$, $B=-10^{-3}$\ and $B=+10^{-1}$, there
is an onset of an ever growing acceleration eventually leading to a
\emph{premature Big Rip}, respective to the one taking place in the $\Lambda
$CDM model. On the other hand, the velocity in the higher order model with
$B=+10^{-3}$\ -- black curve in Fig. \ref{Fig-Behaviour}\ -- reaches a maximum
value at $z\simeq-0.15$, the acceleration then changes the sign, leading to a
positive deceleration parameter. This deceleration is sufficiently intense to
produce a \emph{Rebouncing}, where the universe reaches a maximum size and
then begins to shrink. This happens at $z\simeq-0.32$.

\begin{figure}[ptbh]
\begin{center}
\includegraphics[
natheight=11.500300in,
natwidth=11.535700in,
height=2.9395in,
width=2.949in
]
{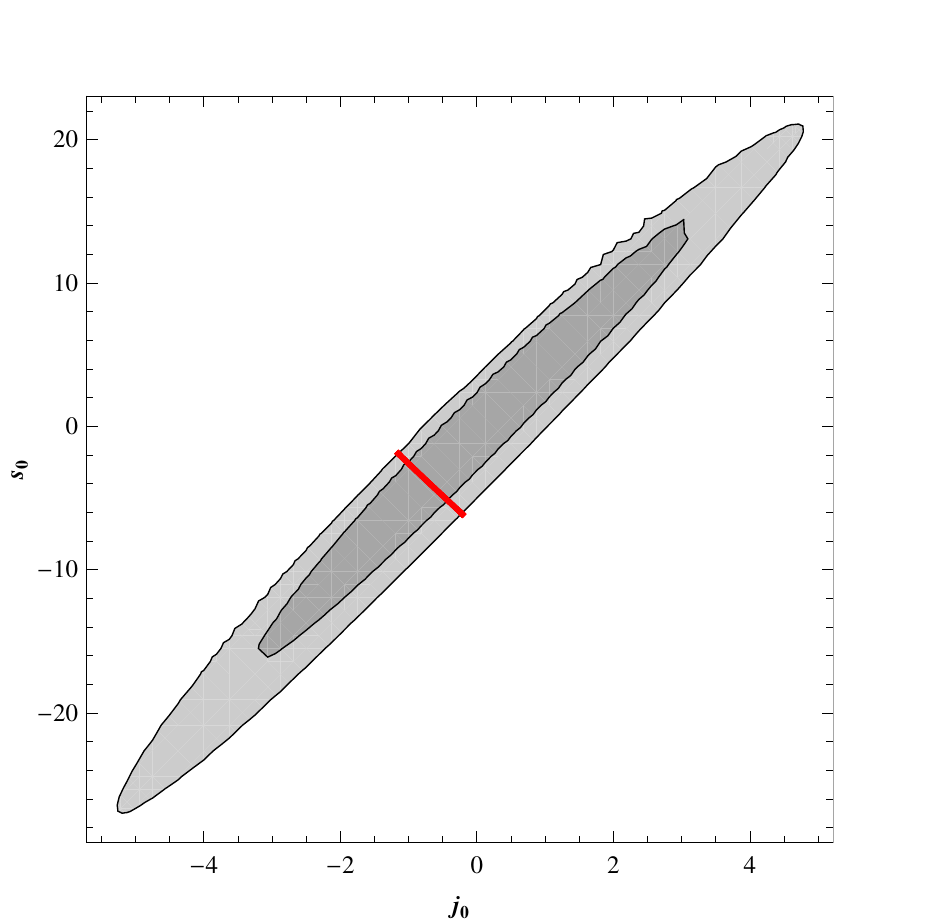} \includegraphics[
natheight=11.333400in,
natwidth=11.368800in,
height=2.9395in,
width=2.949in
]
{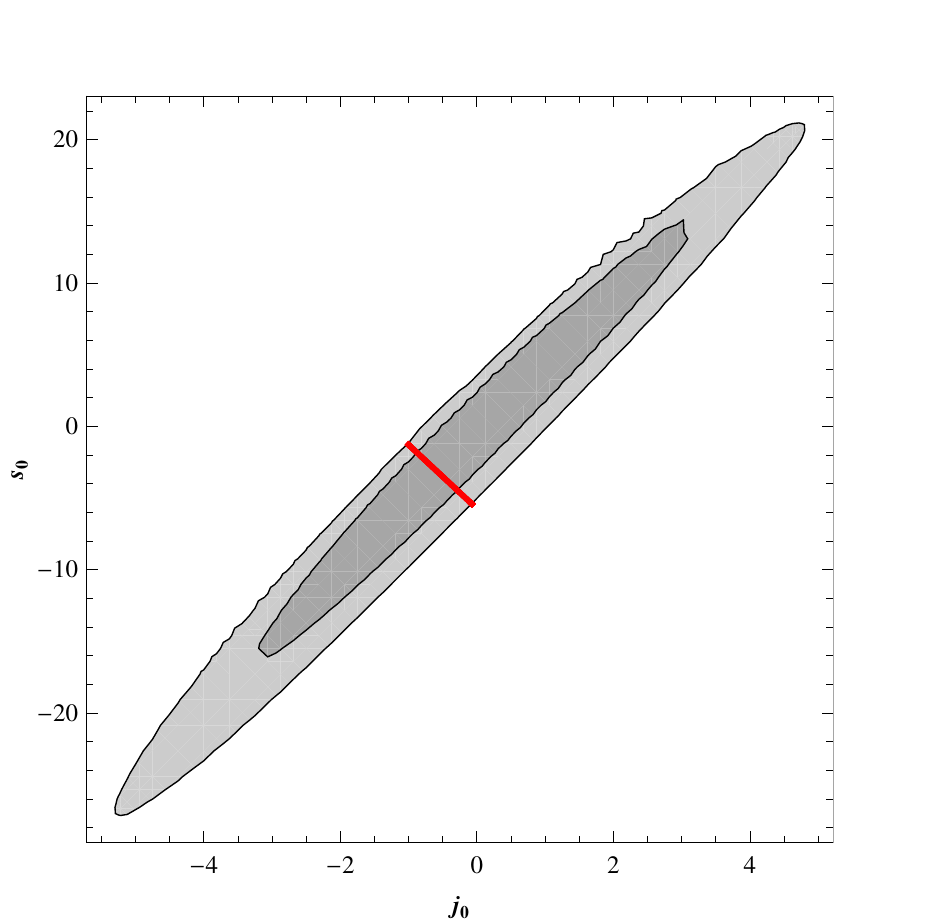}
\end{center}
\caption{$\left(  j_{0},s_{0}\right)  $ confidence regions for $B=-10^{-1}$
(left)\ and $B=+10^{-1}$\ (right) with $q_{0}$\ and $\Omega_{m0}$\ set to
their best-fit values.}%
\label{Fig-JerkSnap}%
\end{figure}

This difference between Big Rip and Rebouncing scenarios deserves a more
careful statistical analysis. We shall consider this in a broader space of
parameters, and not using only the best-fit values for the parameters of our
model, as we have done so far. A careful analysis shows that the deceleration
parameter $q_{0}$\ influences weakly on the possible future regimes of our
model. The parameters that decisively decide in favor of a Big Rip scenario or
a Rebouncing event are the jerk $j_{0}$\ and the snap $s_{0}$. Fig.
\ref{Fig-JerkSnap}\ displays the $\left(  j_{0},s_{0}\right)  $ confidence
regions plots for $B=\pm10^{-1}$.

The straight red lines passing through the confidence regions of the plots
Fig. \ref{Fig-JerkSnap} separate the sector of the space of parameters where
the Rebouncing occurs (lower part) from that region of values corresponding to
a Big Rip (upper part). A numerical analysis shows that 39\% of the $\left(
j_{0},s_{0}\right)  $ confidence region allows for the Rebouncing scenario in
the higher order model with $B=-10^{-1}$ and 42\% of the confidence region for
the model with $B=+10^{-1}$\ leads to Rebouncing. A similar analysis shows
that the model with $B=-10^{-3}$\ produces a premature Big Rip for all the
values of the pair $\left(  j_{0},s_{0}\right)  $ in the 95\% confidence
region; converselly, the model with $B=+10^{-3}$\ engenders Rebouncing within
the 95\% CR.

At last, it is interesting to explore the features of the model for $z\gtrsim1$.
Extrapolating the results presented in Fig. \ref{Fig-Behaviour}\ we verified that
our model \emph{does not reproduce} the decelerated phase
typical of a matter dominated Friedmann universe. Besides, for $10 \lesssim z \lesssim 1000$
the higher order model exhibits a behaviour compatible with\footnote{The result (51) is still valid even when radiation is added to the model.}
\begin{equation}
\frac{1}{3}<\omega<\frac{1.25}{3}\;, \label{w interval}\end{equation}
where the equation of state is $P=\omega\rho$, i.e. the equation for a perfect fluid
in the standard Friedmann cosmology.
On the other hand, a linear perturbative analysis shows that
the perturbations in gravitation and matter fields do not grow for the equation of state with $\omega$ as given in  (\ref{w interval}). Therefore, the model proposed here can not describe the structure formation in the early universe.%


\section{Final Remarks \label{sec-Conclusion}}

In this paper we have studied the phenomenological features of a particular
modified gravity model built with an action integral dependent on the Ricci
scalar $R$ and also on the contraction $\nabla_{\mu}R\nabla^{\mu}R$\ of its
covariant derivatives.\ The Friedmann equations resulting from the field
equation and the FLRW line element involve sixth-order time-derivatives of the
scale factor $a$. We solved those differential equations numerically -- but
without any approximation -- assuming a null cuvature parameter ($\kappa
=0$)\ and pressureless matter. Our higher order model depends on five
parameters, namely $\left[  \mathcal{M}\right]  =\left\{  q_{0},j_{0}%
,s_{0},\Omega_{m0};B\right\}  $. The present day value of the Hubble constant
is taken as given -- and not constrained by data fit -- the reason for that
being consistency with the $f$-gas data analysis. Four of the five parameters
$\left[  \mathcal{M}\right]  $\ are determined by the best-fit to the
data\ for the SN Ia $d_{L}\left(  z\right)  $ and fraction of gas in galaxy
clusters $d_{A}\left(  z\right)  $\ available from Refs.\cite{Union2}\ and
\cite{Allen2008}, respectively.

The shape of the likelihood function $\mathcal{L} $\ in terms of $B$ shows
that it is not possible to determine this parameter by maximization. We were
forced to select some values of $B$ and proceed to the best-fit of the other
parameters in the set $\left[  \mathcal{M}\right]  $. It was clear that the
$\Omega_{m0}$\ is statistically independent of the initial conditions,
regardless the values of $B$ and the other best-fit parameters. In addition
the sign of both $j_{0}$\ and $s_{0}$\ depend on $B$, so that the behaviour of
the model towards the future is not unique.

The confidence regions for the pair of parameters $\left(  q_{0},j_{0}\right)
$, $\left(  q_{0},\Omega_{m0}\right)  $ and $\left(  j_{0},s_{0}\right)  $
were built. They showed that $q_{0}$\ and $j_{0}$\ are strongly correlated
parameters, whereas $q_{0}$\ and $\Omega_{m0}$\ are weakly correlated
quantities. This was confirmed by the results in Table \ref{table-Covariances}%
. The plot of the confidence region for the pair $\left(  j_{0},s_{0}\right)
$\ is relevant to understand the future evolution of our higher-order model,
which can exhibit either a premature Big Rip or a Rebouncing depending on the
value of $B$ and on the values of $j_{0}$\ and $s_{0}$.

If additional data for high $z$ objects were used, as Gamma Ray Bursts, for instance, they
might reduce the uncertainties in our estimations of the jerk and
the snap, and help us to decide if the Rebouncing scenario is favored with
respect to the Big Rip, or conversely.

A positive feature of the particular $f\left(  R,\nabla R\right)  $-model
presented here is its consistency with the standard $\Lambda$CDM dynamics at
recent times. Maybe this is anticipated because the model has a great number
of free parameters to be fit to the data. The high number of kinematic
parameters (i.e., those involving acceleration and its first and second
derivatives) is a consequence of the model being of the fifth-order in
derivatives of $H$, a fact that could always make room for agreement with the
observations. However, the freedom for bias is not unlimited. Even with all
these free parameters there is a constraint to the dynamics of the model
towards the past.

Besides this consistency with the standard $\Lambda$CDM dynamics (at recent times) the model does not reproduce a large scale structure formation phase. This shows that just fitting a model to actual data is not enough to falsify or prove the model. General essential characteristic must be present too, as the instabilities allowing the large scale structure formation. Therefore, the extra
term presented in the action (\ref{Action}) should be ruled out as viable alternative for dark energy.%

\subparagraph{Acknowledgements}

This paper is dedicated to Prof. Mario Novello on the ocasion of his 70th
birthday. RRC thanks FAPEMIG-Brazil (grant CEX--APQ--04440-10) for financial
support. CAMM is grateful to FAPEMIG-Brazil for partial support. LGM
acknowledges FAPERN-Brazil for financial support.


\bigskip

\bigskip


\begin{thebibliography}{99}                                                                                               %


\bibitem {WMAP Collaboration}D. N. Spergel \textit{et al}., First year
Wilkinson Microwave Anisotropy Probe (WMAP) observations: Determination of
cosmological parameters, \textit{Astrophys. J. Suppl.} \textbf{148} (2003),
175; E. Komatsu \textit{et al}., Seven-year Wilkinson Microwave Anisotropy
Probe (WMAP) observations: Cosmological interpretation, \textit{Astrophys. J.
Suppl.} \textbf{192}:18 (2011).

\bibitem {Planck}Planck Collaboration: P. A. R. Ade \textit{et al}., Planck
2013 results. I. Overview of products and scientific results, arXiv: 1303.5062
(2013); Planck Collaboration: P. A. R. Ade \textit{et al}., Planck 2013
results. XVI. Cosmological parameters, arXiv: 1303.5076 (2013).

\bibitem {SDSS Collaboration}D. J. Eisenstein \textit{et al}., Detection of
the Barion Acoustic Peak in the large-scale correlation function of SDSS
luminous red galaxies, \textit{Astrophys. J.} \textbf{633} (2005), 560; W. J.
Percival \textit{et al.}, Baryon Acoustic Oscillations in the Sloan Digital
Sky Survey Data Release 7 Galaxy Sample, \textit{Mon. Not. Roy. Astron. Soc.}
\textbf{401} (2010), 2148.

\bibitem {2dFGRS Collaboration}S. Cole \textit{et al.}, The 2dF Galaxy
Redshift Survey: Power-spectrum analysis of the final dataset and cosmological
implications, \textit{Mon. Not. Roy. Astron. Soc.} \textbf{362} (2005), 505;
F. Beutler, The 6dF Galaxy Survey: Baryon Acoustic Oscilations and the local
Hubble constant, \textit{Mon. Not. Roy. Astron. Soc}. \textbf{416} (2011), 3017.

\bibitem {SNLS Collaboration}P. Astier \textit{et al.}, The Supernova legacy
Survey: Measurement of $\Omega_{M}$, $\Omega_{\Lambda}$\ and $w$\ from the
first year data set, \textit{Astron. Astrophys.} \textbf{447} (2006), 31.

\bibitem {Supernova Search Team Collaboration}A. G. Riess \textit{el al.},
Type Ia Supernova discoveries at $z>1$ from the Hubble Space Telescope:
Evidence for past deceleration and constraints on dark energy evolution,
\textit{Astrophys. J.} \textbf{607} (2004), 665; A. G. Riess \textit{el al.},
New Hubble Space Telescope discoveries of Type Ia Supernovae at $z>1$:
Narrowing constraints on the early behavior of dark energy, \textit{Astrophys.
J.} \textbf{659} (2007), 98;

\bibitem {ESSENCE Collaboration}W. M. Wood-Vasey \textit{et al.},
Observational constraints on the nature of the dark energy: first cosmological
results from the ESSENCE Supernova Survey, \textit{Astrophys. J.} \textbf{666}
(2007), 694.

\bibitem {Union2}R. Amanullah \textit{et al.}, Spectra and Hubble Space
Telescope light curves of six type Ia supervonae at $0.511<z<1.12$\ and the
Union2 Compilation, \textit{Astrophys. J.} \textbf{716} (2010), 712.

\bibitem {Union21}N. Suzuki \textit{et al.}, The Hubble Space Telescope
Cluster Supernova Survey: V. Improving the Dark Energy Constraints Above $z>1$
and Building an Early-Type-Hosted Supernova Sample, \textit{Astrophys. J.}
\textbf{746} (2012), 85.

\bibitem {Schindler2002}S. Schindler, $\Omega_{M}$-different ways to determine
the matter density of the universe, \textit{Space Science Reviews}
\textbf{100} (2002), 299.

\bibitem {Allen2008}S. W. Allen \textit{et al.}, Improved constraints on dark
energy from Chandra X-ray observations of the largest relaxed galaxy clusters,
\textit{Mon. Not. Roy. Astron. Soc.} \textbf{383} (2008), 879.

\bibitem {Weinberg1989}S. Weinberg, The cosmological constant problem,
\textit{Rev. Mod. Phys.} \textbf{61} (1989) 1.

\bibitem {Riess1998}A. G. Riess \textit{et al.}, Observational evidence from
supernovae for an accelerating universe and a cosmological constant,
\textit{Astron. J.} \textbf{116} (1998), 1009.

\bibitem {Perlmutter1999}S. Perlmutter \textit{et al.}, Measurements of
$\Omega$\ and $\Lambda$\ from 42 high-redshift supernovae, \textit{Astrophys.
J.} \textbf{517} (1999), 565.

\bibitem {Amendola2010}L. Amendola e S. Tsujikawa, \textit{Dark Energy: Theory
and Observations}, Cambridge University Press, New York (2010).

\bibitem {Ratra1988}B. Ratra and P. J. E. Peebles, Cosmological consequences
of a rolling homogeneous scalar field, \textit{Phys. Rev D} \textbf{37}
(1988), 3406.

\bibitem {Caldwell1998}R. R. Caldwell, R. Dave, and P. J. Steinhardt,
Cosmological imprint of an energy component with general equation-of-state,
\textit{Phys. Rev Lett.} \textbf{80} (1998), 1582.

\bibitem {Carrol1998}S. M. Carrol, Quintessence and the rest of the world,
\textit{Phys. Rev Lett.} \textbf{81} (1998), 3067.

\bibitem {Hebecker2001}A. Hebecker and C. Wetterich, Natural quintessence?,
\textit{Phys. Lett. B} \textbf{497} (2001), 281.

\bibitem {Gabi2007}L. Amendola, G. C. Campos, and R. Rosenfeld, Consequences
of dark matter-dark energy interaction on cosmological parameters derived from
SN\ Ia data, \textit{Phys. Rev. D} \textbf{75} (2007), 083506.

\bibitem {Chiba2000}T. Chiba, T. Okabe, and M. Yamaguchi, Kinetically driven
quintessence, \textit{Phys. Rev. D} \textbf{62} (2000), 023511.

\bibitem {Armendariz-Picon2001}C. Armendariz-Picon, V. F. Mukhanov, and P. J.
Steinhardt, Essentials of k-essence, \textit{Phys Rev. D} \textbf{63} (2001), 103510.

\bibitem {Kamenshchik2001}A. Y. Kamenshchik, U. Moschella, and V. Pasquier, An
alternative to quintessence, \textit{Phys. Lett. B} \textbf{511} (2001), 265.

\bibitem {Bento2002}M. C. Bento, O. Bertolami, and A. A. Sen, Generalized
Chaplygin gas, accelerated expansion and dark energy-matter unification,
\textit{Phys. Rev. D} \textbf{66} (2002), 043507.

\bibitem {Capozziello2011}S. Capozziello and M. De Laurentis, Extended
theories of gravity, \textit{Phys. Rep.} \textbf{509} (2011), 167.

\bibitem {Odintsov2011}S. Nojiri and S. D. Odintsov, Unified cosmic history in
modified gravity: from $f\left(  R\right)  $ theory to Lorentz non-invariant models,
\textit{Phys. Rep.} \textbf{505} (2011), 59.

\bibitem {Dvali2000}G. R. Dvali, G. Gabadadze, and M. Porrati, 4D gravity on a
brane in 5D minkowski space, \textit{Phys. Lett. B} \textbf{485} (2000), 208.

\bibitem {Sahni2003}V. Sahni and Shtanov, Braneworld models of dark energy,
\textit{JCAP} \textbf{0311} (2003), 014.

\bibitem {Bartolo2000}N. Bartolo and M. Pietroni, Scalar tensor gravity and
quintessence, \textit{Phys. Rev. D} \textbf{61} (2000), 023518.

\bibitem {Perrotta2000}F. Perrota, C. Baccigalupi, and S. Matarrese, Extended
quintessence, \textit{Phys. Rev. D} \textbf{61} (2000), 023507.

\bibitem {FinslerDE}Z. Chang, X. Li, Modified Friedmann model in
Randers-Finsler space of approximate Berwald type as a possible alternative to
dark energy hypothesis, \textit{Phys. Lett. B }\textbf{676} (2009) 173.

\bibitem {LyraDE2011}K. S. Adhav, LRS Bianchi Type-I Universe with Anisotropic
Dark Energy in Lyra Geometry, \textit{Int. J. Astr. Astrophys.}, \textbf{1}
No. 4, (2011) 204. doi: 10.4236/ijaa.2011.14026.

\bibitem {LyraDKP}R. Casana, C. A. M. de Melo, B. M. Pimentel, Massless DKP
field in a Lyra manifold, \textit{Class. Quantum Gravity} \textbf{24} (2007) 723.

\bibitem {Capozziello2002}S. Capozziello, Curvature quintessence, \textit{Int.
J. Mod. Phys. D} \textbf{11} (2002), 483.

\bibitem {DeFelice2010}A. De Felice and S. Tsujikawa, $f\left(  R\right)
$\ theories, \textit{Living Rev. Rel.} \textbf{13} (2010), 3.

\bibitem {Sotiriou2010}T. P. Sotiriou and V. Faraoni, $f\left(  R\right)
$\ theories of gravity,\textit{ Rev. Mod. Phys.} \textbf{82} (2010), 451.

\bibitem {Santos2008}J. Santos \textit{et al.}, Latest supernovae constraints
on $f\left(  R\right)  $ cosmologies, \textit{Phys. Lett. B} \textbf{669}
(2008), 14.

\bibitem {Pires2010}N. Pires, J. Santos and J. S. Alcaniz, Cosmographic
constraints on a class of Palatini $f\left(  R\right)  $ gravity,
\textit{Phys. Rev. D} \textbf{82} (2010), 067302.

\bibitem {Odintsov2006}S. Nojiri and S. D. Odintsov, Introduction to modified
gravity and gravitational alternative for dark energy, 
\textit{Int. J. Geom. Meth. Mod. Phys.} \textbf{4} (2007), 115.

\bibitem {Bengochea2009}G. Bengochea and R. Ferraro, Dark torsion as the
cosmic speed-up, \textit{Phys. Rev. D} \textbf{79} (2009), 124019.

\bibitem {Linder2010}E. V. Linder, Einstein's other gravity and the
acceleration of the universe, \textit{Phys. Rev. D} \textbf{81} (2010), 127301.

\bibitem {Bamba2011}K. Bamba \textit{et al.}, Equation of state for dark
energy in $f\left(  T\right)  $ gravity, \textit{JCAP} \textbf{1101} (2011), 021.

\bibitem {Gottlober1990}S. Gottl\"{o}ber, H-J Schmidt, A. A. Starobinsky,
Sixth-order gravity and conformal transformations, \textit{Class. Quantum
Gravity} \textbf{7} (1990), 893.

\bibitem {Biswas2006}T. Biswas, A. Mazumdar and W. Siegel, Bouncing universes
in string-inspired gravity, \textit{JCAP} \textbf{0603} (2006), 009.

\bibitem {Biswas2010}T. Biswas, T. Koivisto and A. Mazumdar, Towards a
resolution of the cosmological singularity in non-local higher derivative
theories of gravity, \textit{JCAP} \textbf{1011} (2010), 008.

\bibitem {Biswas2012}T. Biswas \textit{et al.}, Stable bounce and inflation in
non-local higher derivative cosmology, \textit{JCAP} \textbf{1208} (2012), 024.

\bibitem {Arkani-Hamed2002}N. Arkani-Hamed \textit{et al.}, Non-local
modification of gravity and the cosmological constant problem, hep-th/0209227.

\bibitem {Barvinsky2003}A. O. Barvinsky, Nonlocal action for long-distance
modifications of gravity theory, \textit{Phys. Lett. B} \textbf{572} (2003), 109.

\bibitem {Annals}R. R. Cuzinatto, C. A. M. de Melo, and P. J. Pompeia, Second
order gauge theory, \textit{Ann. Phys.} \textbf{322} (2007), 1211.

\bibitem {EPJC}R. R. Cuzinatto, C. A. M. de Melo, and P. J. Pompeia, Gauge
formulation for higher order gravity, \textit{Eur. Phys. J.} C \textbf{53}
(2008), 99.

\bibitem {AstrophysSpaceSci}R. R. Cuzinatto, C. A. M. de Melo, and P. J.
Pompeia, Cosmic acceleration from second order gauge gravity,
\textit{Astrophys. Space Sci.} \textbf{332} (2011), 201.

\bibitem {Visser2004}M. Visser, Jerk, snap, and the cosmological equation of
state, \textit{Class.Quant.Grav.} \textbf{21} (2004) 2603.

\bibitem {Medeiros2012}L. G. Medeiros, Realistic Cyclic Magnetic Universe,
\textit{Int. J. Mod. Phys. D} \textbf{21} (2012), 1250073.

\bibitem {Holz2005}D. E. Holz and E. V. Linder, Safety in numbers:
Gravitational Lensing Degradation of the Luminosity Distance-Redshift
Relation, \textit{Astrophys. J.} \textbf{631} (2005) 678.

\bibitem {Kirkman2003}D. Kirkman \textit{et al.}, The cosmological baryon
density from the deuterium to hydrogen ratio towards QSO absorption systems:
D/H towards Q1243+3047, \textit{Astrophys. J. Suppl.} \textbf{149} (2003) 1.

\bibitem {Riess2011}A. G. Riess \textit{el al.}, A 3\% solution: determination
of the Hubble constant with the Hubble Space Telescope and Wide Field Camera
3, \textit{Astrophys. J.} \textbf{730} (2011), 119.

\bibitem {Beringer2012}J. Beringer \textit{et al.}, Review of Particle
Physics, \textit{Phys. Rev. D} \textbf{86} (2012), 010001.
\end{thebibliography}
\end{document}